\documentclass[lettersize,journal]{IEEEtran}
\usepackage{graphicx}
\usepackage{cite}
\usepackage{picinpar}
\usepackage{amsmath}
\usepackage{url}
\usepackage{flushend}
\usepackage[utf8]{inputenc}
\usepackage{colortbl}
\usepackage{soul}
\usepackage{multirow}
\usepackage{pifont}
\usepackage{color}
\usepackage{alltt}
\usepackage[hidelinks]{hyperref}

\providecommand{\arxivurl}[1]{\texttt{\detokenize{#1}}}
\usepackage{enumerate}
\usepackage{siunitx}
\usepackage{pbox}
\usepackage{booktabs}
\usepackage{tabularx}
\usepackage{array}

\providecommand{\IEEEmembership}[1]{\textit{#1}}
\providecommand{\IEEEPARstart}[2]{#1#2}

\begin{document}
\title{Artificial Intelligence for Power-Converter-Rich Electrical Systems: A Review}

\author{
	{Pengfeng Lin,~\IEEEmembership{Senior Member,~IEEE}, 
	Yuan Gao,~\IEEEmembership{Senior Member,~IEEE}, 
	Yuxi Tang,~\IEEEmembership{Student Member,~IEEE},
	Muhammad Waqas Qaisar,~\IEEEmembership{Student Member,~IEEE},
	Peifeng Hui,~\IEEEmembership{Student Member,~IEEE},	
	Chuanlin Zhang,~\IEEEmembership{Senior Member,~IEEE},
	Miao Zhu,~\IEEEmembership{Senior Member,~IEEE},
	Xiaoyong Cao, ~\IEEEmembership{Member,~IEEE},
	Chia-chi Chu, ~\IEEEmembership{Fellow,~IEEE}
	Peng Wang,~\IEEEmembership{Fellow,~IEEE}
	}

	\thanks{
		
		Pengfeng Lin is with University of Cambridge, UK and Shanghai Jiao Tong University, China (Email: linpengfeng@ieee.org).
		
		Yuan Gao is with University of Leicester, UK (Email: yuan.gao@leicester.ac.uk).
		
		Yuxi Tang, Miao Zhu and Xiaoyong Cao are with  Shanghai Jiao Tong University, China (Email: 1804018234@qq.com, miaozhu@sjtu.edu.cn, xiaoyong.cao@sjtu.edu.cn).
		
		Peifeng Hui and Chuanlin Zhang are with Shanghai University of Electric Power, China (Email: peifenghui@mail.shiep.edu.cn, clzhang@shiep.edu.cn).
		
		Chia-chi Chu is with National Tsing Hua University, Taiwan, R.O.C (Email: cchu@ee.nthu.edu.tw)
		
		Peng Wang is with Nanyang Technological University, Singapore (epwang@ntu.edu.sg).
		
		
		
	}
}

\maketitle
	
\begin{abstract}
Power-converter-rich electrical systems, formed by renewable generation, electrified transportation, and inverter-based resources, exhibit strongly nonlinear dynamics, multi-physics design tradeoffs, fast control requirements, and growing reliability and cybersecurity constraints. These characteristics strain workflows that rely only on physics-based modeling, sequential optimization, and rule-based operation. This paper reviews artificial intelligence (AI) for power-converter-rich electrical systems through a life-cycle and deployment-readiness perspective. The literature is organized across converter design, real-time control, system-level operation, and compliance-oriented governance. For design, we examine surrogate modeling, topology and parameter synthesis, EMI/EMC-aware optimization, reliability-oriented design, and knowledge-assisted workflows. For control, we compare supervised learning, reinforcement learning, learning-augmented predictive control, and safety-constrained learning according to their role in closed-loop implementation. For operations, we focus on microgrid coordination, forecasting, distribution-system observability, privacy-preserving coordination, and cyber-resilient operation where converter-interfaced resources shape the operating problem. Across these stages, the review emphasizes deployment-critical gaps, including stability certification, constraint satisfaction, interpretability, extrapolation, data efficiency, sim-to-real transfer, embedded latency, cybersecurity, privacy, and standards alignment. The resulting taxonomy is intended to clarify where AI is already useful as an engineering support tool and where further validation is needed before autonomous or safety-critical deployment.
\end{abstract}


{}

\definecolor{limegreen}{rgb}{0.2, 0.8, 0.2}
\definecolor{forestgreen}{rgb}{0.13, 0.55, 0.13}
\definecolor{greenhtml}{rgb}{0.0, 0.5, 0.0}

\section{Introduction}

\IEEEPARstart{T}{he} rapid penetration of renewable generation, electrified transportation, and inverter-interfaced resources is reshaping modern power systems into power-converter-rich electrical systems \cite{TPEL.2026.3675661,TIA.2025.3529797}. In these low-inertia and increasingly decentralized networks, power electronic converters are no longer auxiliary interfaces. They increasingly define how energy is converted, routed, regulated, protected, and coordinated. As a result, converter design, control, operation, and maintenance have become coupled engineering tasks across device, circuit, controller, and grid levels \cite{TPEL.2020.3024914}. This transition brings substantial complexity because converters combine nonlinear switching behavior, multi-physics coupling, fast electromagnetic transients, thermal stress, aging mechanisms, and strong interactions with uncertain grid conditions. Conventional physics-based workflows, including finite-element analysis, small-signal linearization, empirical degradation modeling, and rule-based operation, remain indispensable. However, they are increasingly strained by high-dimensional design spaces, short control time scales, heterogeneous data sources, and stochastic operating environments \cite{RSER.2025.116591}.

AI provides a complementary paradigm for addressing these computational and operational bottlenecks. ML, DL, RL, FL, GNNs, physics-informed learning, and LLM-assisted engineering tools have been introduced to accelerate converter design, enhance adaptive control, support distributed coordination, and improve condition monitoring \cite{APENERGY.2025.126923}. These methods can extract latent patterns from simulation and measurement data, approximate expensive physical models, learn sequential decision policies, and enable localized or collaborative intelligence in converter-dominated grids. However, applying AI to safety-critical electrical infrastructure is fundamentally different from using AI as an offline prediction tool. Practical deployment must address formal stability, constraint satisfaction, interpretability, extrapolation robustness, data efficiency, real-time hardware latency, cybersecurity, privacy, and compliance with engineering standards.

Although several reviews have summarized AI applications in power electronics and power systems, important gaps remain. Existing studies, such as \cite{TPEL.2020.3024914} and \cite{APENERGY.2025.126923}, provide valuable surveys of algorithmic progress. However, they do not fully connect AI methods with the engineering life cycle of power-converter-rich electrical systems. In particular, three limitations motivate this review. First, many prior works emphasize performance metrics, such as prediction accuracy or control reward, while giving less attention to deployment barriers including Lyapunov stability certification, safe exploration, embedded-controller latency, and sim-to-real transfer. Second, existing classifications often group methods by algorithm family rather than by converter life-cycle function, making it difficult to identify where AI is most mature and where engineering validation is still weak. Third, cross-cutting issues such as physics consistency, data governance, cyber-physical security, and AI compliance standards are often treated separately, even though they jointly determine whether AI can be trusted in converter-dominated infrastructure.

\subsection{Survey Scope and Classification Protocol}

\begin{figure*}[htbp]
\vspace{-1.0em}
\centering
\includegraphics[width=1\linewidth]{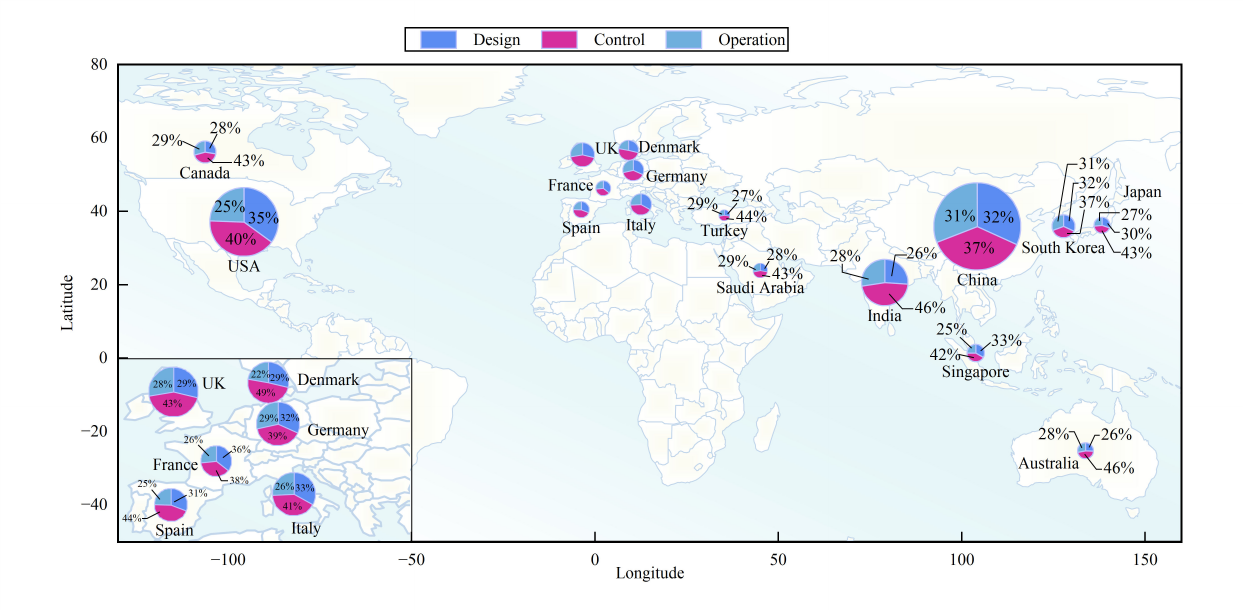}
\caption{Global research landscape of AI applications in power-converter-rich electrical systems within the surveyed literature set.}
\label{Piemap}
\vspace{-0.8em}
\end{figure*}

\begin{figure}[htbp]
\vspace{-1.0em}
\centering
\includegraphics[width=1.0\linewidth]{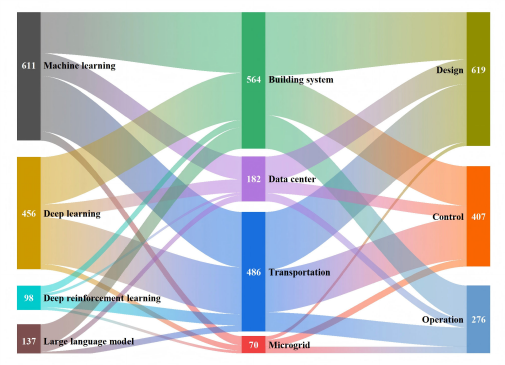}
\caption{Distribution of AI paradigm adoption across application domains and life-cycle phases in the surveyed power-converter-rich electrical-systems literature.}
\label{Sankey}
\vspace{-0.8em}
\end{figure}

This review uses the cited literature set as a scoped map rather than as an exhaustive bibliometric census of all AI-in-energy publications. Each work was screened according to three questions: whether an AI method is central to the contribution, whether the target system involves power converters, inverter-interfaced resources, microgrids, distribution grids, or converter-dominated operation, and which engineering life-cycle role is most directly supported by the evidence. Papers were then assigned to the primary categories of design, control, operation, and compliance-oriented governance. When a work spanned multiple stages, the primary category was determined by the main validation task rather than by the algorithm name alone. Figures \ref{Piemap} and \ref{Sankey} therefore summarize the structure of the surveyed corpus and should be interpreted as a scoped research landscape, not as a database-wide publication count.

Within this surveyed set, Fig.~\ref{Piemap} summarizes the geographical distribution of AI-related research priorities across the major life-cycle phases of converter-rich electrical systems. The overall distribution of AI paradigms and application topics is shown in Fig. \ref{Sankey}. China and the United States contribute a large share of the surveyed work and cover design, control, and operation broadly. From the perspective of regional emphasis, control-oriented studies occupy a large proportion in China and India, whereas design-oriented research is more visible in the United States, South Korea, and several European countries, including the UK, Germany, and Italy. Operation-related studies maintain a relatively balanced share across most regions. These patterns should be read as qualitative indicators of the reviewed corpus rather than as normalized global research-output statistics.

This paper addresses the above gaps by reviewing AI-enabled methods from a full life-cycle perspective. For converter design, AI is used to accelerate multi-objective optimization under power density, efficiency, electromagnetic interference (EMI), thermal, cost, and reliability constraints. High-fidelity finite-element analysis (FEA) and circuit simulation provide accurate evaluation, but their computational burden can delay iterative design \cite{CSEEJPES.2020.02700}. ML-based surrogate models, including deep neural networks (DNNs), convolutional neural networks (CNNs), and other regression architectures, can approximate nonlinear mappings between geometry, material, layout, and multi-physics responses \cite{JESTIE.2022.3198504}. RL-based frameworks further extend AI from performance prediction to automated topology and parameter synthesis \cite{ICCAD.2021.9643548}. Nevertheless, design-stage AI remains limited by data requirements, weak extrapolation outside the training distribution, insufficient uncertainty quantification, and the need for physics-based verification \cite{TPWRS.2022.3162473}.

For converter control, AI is increasingly investigated as a means of handling nonlinear dynamics, uncertain parameters, changing network impedance, and decentralized grid operation. Deep reinforcement learning (DRL) agents, including methods based on deep deterministic policy gradient and soft actor-critic, have been explored for switching optimization, zero-voltage-switching range extension, transient regulation, and adaptive converter control \cite{TIA.2025.3626472,OJPEL.2025.3619673}. Compared with fixed-gain proportional-integral control and finite-control-set model predictive control, learning-based control can offer improved adaptability under complex operating conditions \cite{TPEL.2024.3358912}. However, its industrial adoption is constrained by the absence of formal stability guarantees, the difficulty of guaranteeing safe behavior during learning, the sim-to-real gap, and the microsecond-level latency requirements of DSP- and FPGA-based controllers \cite{OJIA.2023.3338534,MIE.2022.3211125}. Physics-informed neural networks (PINNs), safety shields, constrained learning, and learning-augmented predictive control are therefore becoming important directions for combining data-driven adaptability with physical consistency and certifiable constraints \cite{TPWRS.2022.3162473,TPWRS.2020.3001919}.

At the system-operation level, converter-rich grids require distributed awareness, low-latency decision-making, privacy-preserving coordination, and cyber-resilient operation. Edge-AI architectures support localized forecasting and control for renewable generation and distributed energy resources \cite{TII.2022.3163137}. FL and GNNs enable networked converters and microgrids to collaboratively train coordination, forecasting, and cyber-attack detection models without directly sharing raw local data \cite{TSG.2024.3466768}. Multi-agent DRL and physics-informed reward structures have also been used for energy management in multi-microgrid systems under uncertainty \cite{TNNLS.2022.3232630}. Such methods are particularly relevant when extreme events, such as cold waves, require robust reserve scheduling and coordinated operation across coupled energy infrastructures \cite{APENERGY.2025.126900}. However, distributed intelligence also expands the attack surface. Communication dependence, compromised edge devices, model poisoning, and adversarial manipulation can undermine the reliability of converter-dominated systems \cite{ACCESS.2021.3131502}.

AI is also becoming important for condition monitoring, fault diagnosis, and predictive maintenance of converter hardware. Semiconductor modules, capacitors, magnetic components, solder layers, bond wires, substrates, and thermal interfaces age under coupled electrical, thermal, and mechanical stress. Purely data-driven ML models can identify degradation signatures from voltage, current, temperature, vibration, and thermal data, but their performance often depends on large labeled datasets and representative run-to-failure records. To alleviate data scarcity, generative adversarial networks (GANs), transfer learning, and domain adaptation have been used to synthesize or transfer fault information across operating conditions \cite{TDSC.2021.3118636}. More fundamentally, physics-informed machine learning (PIML) embeds converter dynamics, thermal models, and degradation laws into the learning process, thereby improving physical plausibility for remaining useful life (RUL) and parameter estimation \cite{APEC.2022.9773482}. Despite these advances, accurate modeling of multi-layer packaging, coupled degradation mechanisms, and high-frequency EMI-contaminated diagnostic signals remains challenging. Therefore, trustworthy AI for operation and maintenance must integrate physical models, uncertainty estimation, robust sensing, and cyber-physical defense \cite{ACCESS.2023.3248511,JAS.2023.123657}.

The main contributions of this review are summarized as follows. First, a life-cycle-oriented taxonomy is established to organize AI applications in power-converter-rich electrical systems across design, control, operation, and compliance-oriented governance. Second, representative AI paradigms are compared according to their engineering roles, including surrogate modeling, topology synthesis, learning-based control, distributed coordination, forecasting, fault diagnosis, and predictive maintenance. Third, deployment-critical limitations are analyzed with emphasis on stability certification, interpretability, data efficiency, hardware latency, sim-to-real transfer, cybersecurity, privacy, and standards alignment. Finally, emerging research directions are discussed, including physics-integrated learning, uncertainty-aware AI, privacy-preserving distributed intelligence, safety-constrained control, and auditable AI workflows.

The remainder of this paper is organized as follows. Section II introduces the AI preliminaries required for the subsequent review. Section III reviews AI-enabled converter design, including surrogate modeling, topology synthesis, EMI-aware design, PCB layout automation, reliability-oriented optimization, and knowledge-driven design assistance. Section IV discusses AI for converter control, with emphasis on learning-assisted predictive control, DRL, safe control, and stability-aware deployment. Section V reviews AI for operation, including energy management, forecasting, resilience, condition monitoring, and distributed digital technologies. Section VI discusses AI compliance standards and trustworthy deployment requirements. Section VII concludes the paper and outlines future research opportunities.

\section{Artificial Intelligence Preliminaries}

This section introduces only the AI concepts needed to interpret the application-oriented review that follows. The goal is not to provide a general AI tutorial, but to clarify how different model families map to converter-rich design, control, and operation tasks.

\subsection{Machine Learning}

Machine learning (ML) provides the algorithmic foundation for data-driven analytics in modern power electronics. By extracting statistical patterns from operational, simulated, and experimentally labeled data, ML techniques can support fault diagnosis, state evaluation, degradation assessment, and performance prediction without relying exclusively on high-dimensional physics-based analytical models. In power-converter-rich electrical systems, the main value of ML lies in its ability to transform measured electrical, thermal, and operational features into actionable diagnostic or decision-support information. The primary paradigms and representative algorithms are summarized below.

\textbf{Supervised Learning:} Supervised learning uses labeled datasets to establish mappings between input features and predefined targets, such as fault categories, health states, degradation levels, or performance indicators. In power electronics, this paradigm is particularly important for regression and classification tasks where operating records or accelerated aging experiments provide known labels. For instance, data-driven prognostic models have been widely investigated for estimating the remaining useful life (RUL) of insulated-gate bipolar transistors (IGBTs), where degradation trajectories provide the basis for supervised health-state assessment \cite{fang2018review}.

\textbf{Unsupervised Learning:} Unsupervised learning aims to uncover latent structures from unlabeled data and is therefore useful when fault labels are scarce, expensive, or practically impossible to obtain. This is especially relevant to large-scale converter networks, where high-frequency measurements are continuously generated but only a small fraction of abnormal events can be manually annotated. Scalable online anomaly detection methods for power electronics networks illustrate this role by identifying abnormal patterns from streaming operational data without requiring exhaustive fault labeling \cite{yu2022congo}.

\textbf{Support Vector Machine (SVM):} SVM constructs maximum-margin decision boundaries in transformed feature spaces and is particularly suitable for classification tasks with limited training samples. In power-electronic diagnostics, SVM-based classifiers have been used to identify fault categories from engineered electrical features, such as wavelet-domain signatures extracted from converter voltage or current signals \cite{cui2017svm}. Compared with the deep architectures discussed in the next subsection, SVM usually requires more explicit feature extraction and kernel selection, but offers relatively low computational complexity and clear decision boundaries for small-sample diagnostic problems.

\textbf{Extreme Learning Machine (ELM):} ELM is a lightweight shallow-learning model based on a single-hidden-layer feedforward structure. By randomly assigning hidden-layer parameters and analytically computing the output weights, ELM avoids the iterative tuning process required by conventional neural networks, which makes it attractive for fast approximation and classification tasks. In power-system applications, transfer ELM frameworks have been used for cross-fault and cross-scale transient stability assessment with limited guide instances, demonstrating the potential of ELM-type models under data-scarce conditions \cite{ren2024transfer}. Nevertheless, its shallow structure limits its ability to automatically learn hierarchical representations from raw high-dimensional measurements.

\textbf{Decision Tree (DT):} DT models construct hierarchical rule-based decision structures and provide strong interpretability compared with many black-box learning methods. Their primary appeal in safety-critical electrical systems lies in the transparency of the extracted decision rules, which allows operators to audit the reasoning behind classification or security-assessment outcomes. Decision-tree-based online voltage security assessment using phasor measurement unit (PMU) data is a representative example of how rule-based learning can support real-time operational decision-making while retaining human interpretability \cite{diao2009decision}.

\textbf{Shallow Neural Network (SNN):} Shallow neural networks, typically comprising one or two hidden layers, serve as basic nonlinear function approximators for electrical dynamics and control-oriented mappings. In converter-related applications, such networks can be used to approximate nonlinear relationships between operating conditions and control references, including maximum power point tracking (MPPT) under rapidly varying irradiance or partial shading conditions \cite{patel2008maximum}. However, because SNNs have limited representation depth, their performance often depends on carefully selected inputs and sufficient operating-condition coverage. This limitation motivates the transition toward deep learning architectures that can learn richer hierarchical features directly from high-dimensional electrical data.

\subsection{Deep Learning}

Deep learning has become a central methodology for modeling high-dimensional and strongly nonlinear electrical data because it can learn hierarchical feature representations directly from raw measurements, thereby reducing the dependence on handcrafted features. In power-converter-rich electrical systems, this capability is particularly valuable, since the available data are rarely independent or homogeneous; instead, they are usually coupled across time, operating conditions, control modes, and network structure. Consequently, the significance of deep learning in this context lies not simply in increasing model depth, but in selecting architectures whose inductive biases are consistent with the structure of the underlying data and the objective of the task \cite{CSEEJPES.2020.02700}.

For temporally ordered measurements, recurrent architectures remain a natural starting point. Long short-term memory (LSTM) networks were originally developed to alleviate the vanishing-gradient problem in conventional recurrent neural networks, thereby enabling the extraction of long-range temporal dependencies from sequential data \cite{hochreiter1997lstm}. Gated recurrent units (GRUs) retain the essential gating mechanism of LSTM while adopting a more compact structure, which often leads to improved training efficiency and lower computational burden \cite{cho2014gru}. In converter- and power-system-related problems, such sequence models are particularly suitable for tasks in which informative patterns are embedded in temporal evolution, including fault diagnosis, state tracking, degradation assessment, and time-series forecasting. Their practical relevance has already been demonstrated in fault diagnosis for DFIG-based wind turbine systems \cite{xue2020lstmfault}.

When the dominant information is concentrated in local patterns rather than long temporal memory, convolutional neural networks (CNNs) are often more effective. Through local receptive fields and weight sharing, CNNs can efficiently extract robust features from waveforms, spectrograms, thermal maps, and trajectory images with relatively high parameter efficiency \cite{krizhevsky2012cnn}. This property makes them especially attractive for converter fault diagnosis and condition monitoring, where subtle local distortions in current or voltage signatures often carry the key physical evidence of abnormal operation. Existing studies further indicate that wavelet-enhanced CNN frameworks can maintain strong diagnostic performance even under limited-sample conditions, underscoring the practicality of convolution-based feature extraction for electrical pattern recognition \cite{hang2023cnnfault}.

More recently, Transformer-based models have broadened the scope of deep sequence modeling by replacing recurrence with attention \cite{vaswani2017transformer}. Their main advantage lies in the direct representation of long-range dependencies together with strong computational parallelism, which is attractive for long-horizon forecasting and multivariable operational analytics. Nevertheless, the adoption of Transformer architectures in power-electronic and power-system problems should be guided by the scale and dependency range of the task, since attention-based models are often more data-intensive and computationally demanding than recurrent or convolutional alternatives \cite{CSEEJPES.2020.02700}.

From the perspective of this review, deep learning in converter-rich electrical systems can therefore be organized into a small number of architecture families associated with different data structures: LSTM and GRU for temporally correlated sequential data, CNN for locally structured waveform- or image-like data, and Transformer for long-range sequence modeling. Graph neural networks provide an important extension for non-Euclidean networked data \cite{liao2022gnn}, and their role becomes increasingly prominent when the analysis moves from component-level intelligence to system-level coordination and topology-aware inference \cite{CSEEJPES.2020.02700}.

\subsection{Deep Reinforcement Learning}

Deep reinforcement learning (DRL) can be viewed as a natural extension of the deep learning paradigm from representation learning to sequential decision-making. While deep learning is primarily concerned with extracting nonlinear mappings from high-dimensional data, reinforcement learning focuses on how an agent interacts with an environment and improves its policy according to long-term return \cite{sutton2018rlintro}. Their combination makes it possible to address control problems in which the system dynamics are complex, strongly coupled, and difficult to model accurately using fixed analytical formulations. This characteristic is particularly relevant to power-converter-rich electrical systems, where uncertainty, nonlinearity, and time-varying operating conditions are intrinsic features rather than exceptions \cite{ye2026rloverview}. Recent overviews of AI-enhanced control in converter-based systems further confirm the growing importance of this direction \cite{gao2023aicontrol}.

From the viewpoint of methodology, DRL differs from conventional deep learning in that the network is no longer used only for classification, regression, or feature extraction, but also for approximating value functions and control policies in closed-loop decision processes \cite{sutton2018rlintro}. This distinction is important in the context of power electronic control, because the objective is not merely to identify system states from data, but to generate control actions that optimize dynamic performance over time \cite{gao2023aicontrol}.

Among representative DRL algorithms, Deep Q-Network (DQN) is the canonical value-based formulation. By approximating the action-value function with a deep neural network, DQN established the basic paradigm of combining reinforcement learning with deep function approximation in high-dimensional settings \cite{mnih2015dqn}. However, because DQN is inherently designed for discrete action spaces, it is more naturally suited to switching-state selection, mode scheduling, or other discretized decision problems than to the direct generation of continuous control signals \cite{ye2026rloverview}.

For continuous-control tasks, Deep Deterministic Policy Gradient (DDPG) provides a more suitable extension. DDPG combines deterministic policy gradients with the actor--critic framework, thereby enabling direct optimization in continuous action spaces \cite{lillicrap2015ddpg}. This property is particularly attractive in converter control problems, where the manipulated variables, such as duty cycles, modulation commands, or reference signals, are intrinsically continuous \cite{ye2026rloverview}. 

Soft Actor--Critic (SAC) further develops the actor--critic framework by incorporating the maximum-entropy principle into policy learning \cite{haarnoja2018sac}. In addition to maximizing expected return, SAC explicitly encourages stochastic exploration, which generally leads to improved robustness and more stable training behavior \cite{haarnoja2018sac}. For nonlinear electrical control problems with demanding convergence requirements, SAC has therefore become one of the most promising DRL formulations in recent power-electronic studies \cite{ye2026rloverview}.

From the perspective of this review, the importance of DRL lies not only in its model-free nature, but also in its ability to learn control-oriented decision policies directly from interaction data \cite{gao2023aicontrol}. At the same time, algorithm selection must remain consistent with the underlying control structure: DQN is more naturally aligned with discrete decision processes \cite{mnih2015dqn}, whereas DDPG and SAC are generally better matched to continuous-control settings \cite{lillicrap2015ddpg,haarnoja2018sac}. For this reason, the latter two are often more relevant to power converter regulation and related real-time control problems \cite{ye2026rloverview,gao2023aicontrol}.

It should also be noted that DRL in this area is no longer limited to conceptual algorithm studies. In practice, simulation environments such as MATLAB/Simulink have provided increasingly accessible workflows for agent construction, training, and closed-loop verification, which lowers the barrier for integrating DRL into simulation-driven controller development.

\subsection{Large Language Models}

Large language models (LLMs) represent a further extension of the deep learning paradigm from task-specific representation learning to large-scale generative modeling of language. Architecturally, modern LLMs are predominantly built on the Transformer, whose attention-based design enables effective modeling of long-range dependencies while retaining strong parallelizability during training \cite{vaswani2017transformer}. On this basis, subsequent studies showed that, as model and data scale increase, language models can exhibit substantial cross-task transfer capability and perform a wide range of tasks through prompting rather than task-specific redesign \cite{brown2020gpt3}.

The significance of LLMs, however, lies not only in scale. An equally important development is the emergence of unified task formulations, in which diverse language processing problems can be cast into a common text-to-text interface \cite{raffel2020t5}. This abstraction substantially broadens the role of the model: instead of serving only as a classifier, regressor, or sequence predictor for a narrowly defined task, the model becomes a general-purpose language interface capable of summarization, explanation, information extraction, code generation, and interactive reasoning under a common prompting framework.

For the purpose of this review, LLMs are particularly relevant because they introduce a new form of AI capability that differs from both conventional machine learning and the control-oriented paradigms discussed earlier. Their strength lies in processing unstructured knowledge, instructions, engineering documents, operational logs, and natural-language queries in a unified manner. This feature makes them especially valuable in settings where the bottleneck is not only numerical prediction or closed-loop regulation, but also knowledge access, human--machine interaction, and cross-domain information integration.

Recent progress has further shifted LLMs from raw next-token predictors toward instruction-following models that better align with user intent. Instruction tuning and feedback-driven alignment have improved their usability in interactive settings, making model outputs more consistent with human expectations and task requirements \cite{ouyang2022instructgpt}. At the same time, purely parametric language models still face clear limitations in factual grounding, provenance, and knowledge updating. Retrieval-augmented generation addresses part of this problem by coupling parametric generation with external non-parametric memory, thereby improving factual support and reducing the dependence on static internalized knowledge alone \cite{lewis2020rag}.

From the perspective of the present review, LLMs should therefore be understood neither as a replacement for physics-based modeling nor as a direct substitute for dedicated control algorithms. Rather, they constitute a distinct methodological layer for language-centered interaction, knowledge organization, and decision support. This positioning is also important for the overall structure of this paper: the present subsection introduces the basic paradigm of LLMs, whereas the following subsection on agentic artificial intelligence will focus on how such models can be embedded into more autonomous task-execution frameworks.

\subsection{Agentic Artificial Intelligence}

Agentic artificial intelligence can be regarded as a further development of the large language model paradigm from language generation to goal-directed task execution. While a standalone large language model mainly serves as a powerful interface for understanding and generating language, an agentic system augments this capability with additional mechanisms such as task decomposition, planning, memory, tool invocation, and interaction with external environments \cite{abouali2026agenticsurvey}. In this sense, agentic AI is not defined by model scale alone, but by the extent to which the system can autonomously organize intermediate steps and adapt its behavior toward a specified objective.

This distinction is particularly important in the context of engineering intelligence. In many practical scenarios, the challenge is not merely to produce an answer in natural language, but to coordinate heterogeneous resources, including simulation tools, optimization modules, databases, monitoring records, and human instructions, within a structured workflow. Agentic AI is therefore more appropriately understood as an orchestration layer built on top of foundation models, rather than as a replacement for domain models or dedicated control algorithms \cite{abouali2026agenticsurvey}.

A central step toward such systems is the integration of reasoning and acting. ReAct showed that language models can interleave verbal reasoning traces with task-specific actions, allowing the model to update its plan while interacting with external information sources or environments \cite{yao2023react}. Toolformer further demonstrated that language models can learn when to call external tools, which tool to invoke, and how to incorporate returned results into subsequent generation \cite{schick2023toolformer}. These developments are significant because they shift the role of the model from passive text generation toward active problem solving.

Another essential ingredient of agentic behavior is the ability to maintain and exploit internal state over time. Generative Agents highlighted the importance of memory, reflection, and planning for producing coherent long-horizon behavior, showing that autonomous agents become substantially more capable when they can store past experience, abstract higher-level reflections, and retrieve relevant context for future actions \cite{park2023generativeagents}. This line of work suggests that agentic intelligence depends not only on instantaneous reasoning ability, but also on the structured management of experience across extended tasks.

Beyond single-agent settings, recent studies have shown that complex tasks can often be handled more effectively through interactions among multiple specialized agents. AutoGen, for example, formalized multi-agent conversation as a practical framework for building systems in which distinct agents collaborate, critique, and refine intermediate outputs in order to accomplish more complex objectives \cite{wu2024autogen}. Such a perspective is especially relevant when problem solving requires multiple competencies that are difficult to concentrate within a single model instance.

From the perspective of this review, agentic artificial intelligence should therefore be viewed as a methodological layer for autonomous workflow organization rather than as an isolated algorithmic category. This positioning is also important for the structure of the paper. The previous subsection introduced large language models as general-purpose language interfaces, whereas the present subsection emphasizes how those models can be embedded into more autonomous systems capable of planning, acting, coordinating tools, and interacting with external environments. The implications of this capability for design, control, and operation will be discussed in the subsequent application-oriented sections.

\section{Artificial Intelligence for Design}

The design of power converters in converter-rich electrical systems has moved beyond the classical task of sizing a known circuit around a fixed specification. It is now a coupled, multi-physics, and multi-objective decision problem in which efficiency, power density, electromagnetic compatibility (EMC), thermal stress, lifetime, cost, manufacturability, and grid-code compliance must be negotiated simultaneously. Analytical design rules, FEA, circuit simulation, and prototype testing remain the technical backbone of this process. However, their sequential and computation-intensive nature limits how many topologies, layouts, components, and mission-profile scenarios can be examined within a realistic design cycle. AI is therefore best understood not as a substitute for physics-based engineering, but as an enabling layer that compresses simulation cost, searches combinatorial design spaces, organizes heterogeneous design knowledge, and closes the loop between specification, synthesis, verification, and documentation.

\begin{figure}[!t]
    \vspace{-1.0em}
    \centering
    \includegraphics[width=1.0\linewidth]{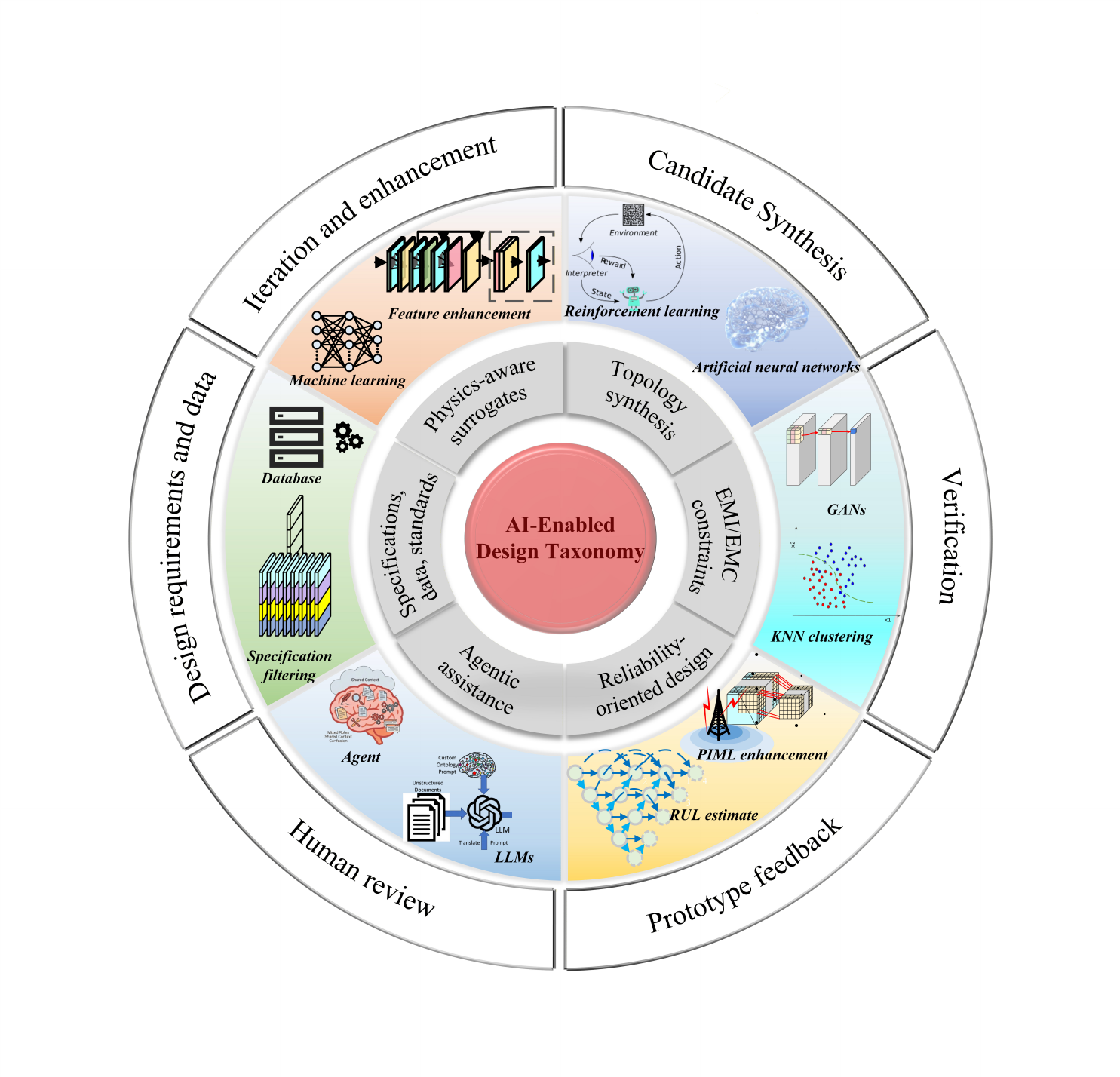}
	\caption{Taxonomy of AI-enabled design for power-converter-rich electrical systems.}
    \label{fig:ai_design_taxonomy}
    \vspace{-0.8em}
\end{figure}

Fig. \ref{fig:ai_design_taxonomy} summarizes this design role from a review perspective. The central shift is from isolated point tools toward an integrated design environment in which AI models propose candidates, physics-based solvers test them, uncertainty estimates determine whether additional evidence is required, and human engineers retain authority over acceptance. On this basis, the literature can be organized into six tightly coupled functions: physics-aware surrogate modeling, topology and parameter synthesis, EMC- and EMI-constrained design, PCB and circuit automation, reliability-oriented optimization, and knowledge-driven or agentic design assistance.

\subsection{Physics-Aware Surrogate Modeling}

Surrogate modeling is the most mature entry point for AI in converter design. The underlying objective is to replace a repeatedly evaluated, expensive mapping from geometry, material properties, layout parasitics, and operating conditions to electromagnetic, thermal, mechanical, loss, or efficiency responses with a fast approximation that can be queried during exploration. DNNs, CNNs, and GANs have been used to predict field distributions and device-level performance in milliseconds, making design-space screening feasible where repeated FEA or coupled multi-physics simulations would be prohibitive \cite{JESTIE.2022.3198504,TPWRS.2022.3162473}. Within the sampled design domain, such surrogates can reduce computation time by several orders of magnitude while preserving sufficient accuracy for early-stage comparison and optimization \cite{TPEL.2020.3024914}.

The key technical issue is that a fast surrogate is useful only if its predictions remain physically credible under the variations imposed by the design task. Purely data-driven models usually require large, well-covered training sets; complex converter structures, including planar transformers and wide-bandgap semiconductor modules, may need thousands of FEA samples before the model becomes reliable \cite{CSEEJPES.2020.02700}. More importantly, a neural surrogate may interpolate accurately but extrapolate with unjustified confidence, producing infeasible geometries, optimistic temperature estimates, or misleading loss predictions outside the training distribution. PINNs and related PIML formulations address this weakness by embedding Maxwell equations, heat-diffusion equations, conservation laws, boundary conditions, or material constraints into the loss function or model architecture \cite{TPWRS.2022.3162473}. In magnetic design, for example, PINN-based formulations have been applied to magnetostatic problems on axisymmetric transformer geometries, allowing magnetic vector potential, inductance, and coupling coefficients to be inferred in a single forward pass while retaining consistency with the governing field equations \cite{FraunhoferPINN}.

The next development is uncertainty-aware surrogate modeling. Instead of returning only a point estimate, the model should indicate whether a candidate is inside a trusted region, whether it violates feasibility constraints, and whether additional high-fidelity simulation is required. Recent stochastic surrogate frameworks combine classifiers that reject infeasible parameter configurations with probabilistic predictors of efficiency, temperature, and uncertainty before heuristic optimizers search for thermally admissible, high-efficiency solutions \cite{Chen2025Probabilistic}. AI-based thermal-resistance surrogates for power-module chip-area optimization have also reported approximately 99.93\% accuracy with far lower computational burden than FEA-based workflows \cite{PowerModuleAI}. For magnetic components, artificial neural networks (ANNs), eXtreme Gradient Boosting (XGBoost), and random forests have been used to estimate inductance and electrodynamic force from dimensional variables, accelerating iterations for air-cored inductors and related structures \cite{Viarouge2024}. These examples show that the strongest role of surrogate models is not autonomous design approval, but risk-aware preselection inside a loop that remains anchored by physics-based simulation and experimental validation.

\subsection{Topology and Parameter Synthesis}

Once fast evaluation models are available, AI can also change the search process itself. Converter design contains both discrete choices, such as topology, semiconductor technology, magnetic core family, and filter structure, and continuous choices, such as duty ratio range, switching frequency, winding geometry, heat-sink size, and control-oriented component values. Exhaustive enumeration is rarely possible because the search space is mixed, constrained, and strongly nonconvex. RL, DRL, evolutionary algorithms, Bayesian optimization, and other metaheuristics have therefore been introduced to explore candidate topologies, component selections, and parameter sets. Fan et al. \cite{ICCAD.2021.9643548} demonstrated an RL-based framework that maps input/output voltage and power requirements to converter topologies by allowing an agent to assemble circuit structures and receive rewards from simulated performance. This line of work moves design from choosing among predefined converter families toward algorithmic generation of candidate architectures from specifications.

For continuous and mixed-variable tuning, ML-assisted multi-objective frameworks are used to approximate Pareto tradeoffs among efficiency, volume, thermal limits, cost, and power density, especially in wide-bandgap-based converters \cite{Lin2025}. ANN surrogates can estimate Pareto fronts with far fewer simulations than brute-force sweeps. Wang et al. \cite{Wang2022} proposed an ANN-based multi-objective methodology for wide-bandgap power electronic converters that reduced computational time by up to 78\% and 67\% relative to numerical modeling and geometric programming, respectively. In their reported 1 kW single-phase inverter prototype, the measured efficiency reached 98.4\% with a power density of 4.57 kW/dm$^{3}$, and experimental errors were below 1\%. Such results are important because they connect algorithmic acceleration with a verified hardware outcome, rather than reporting optimization metrics alone.

Magnetic components are a particularly demanding synthesis target because their performance depends simultaneously on geometry, winding arrangement, core loss, leakage flux, thermal behavior, insulation spacing, and manufacturability. Bayesian optimization, metaheuristics, and deep learning have been applied to automate the shape and winding design of high-frequency transformers and inductors \cite{Shen2024}. Shen et al. \cite{Shen2024} reviewed expert systems, fuzzy logic, metaheuristics, and ML for loss estimation, configuration optimization, and automated design of high-frequency magnetic components. The remaining bottleneck is verification rather than search alone: every topology or parameter set generated by an optimizer must still be traceable to circuit constraints, device ratings, small-signal and transient margins, EMC limits, thermal limits, and practical layout rules before it can be accepted as an engineering solution.

\subsection{EMC- and EMI-Constrained Design}

EMC is one of the design constraints most strongly amplified by wide-bandgap devices. SiC and GaN switches enable high efficiency and high power density, but their fast voltage and current transitions excite parasitic capacitances and inductances across a broad frequency range, from 150 kHz to beyond 100 MHz. As a result, EMI becomes highly sensitive to package parasitics, gate-drive settings, modulation, PCB layout, grounding, shielding, cable configuration, and filter design \cite{Li2026}. Conventional EMI design is commonly performed through measurement, prototype iteration, and empirical filter tuning. This approach remains necessary for compliance, but it becomes increasingly expensive when switching frequency, power density, and regulatory margins all increase.

AI-based EMI design can be divided into three roles: spectral prediction, inverse filter or layout guidance, and adaptive mitigation. Tlig et al. \cite{Tlig2026} compared ANNs, recurrent neural networks (RNNs), k-nearest neighbor (KNN), random forests, and particle swarm optimization for conducted-EMI prediction in DC-DC converters. The KNN model achieved $R^{2}$ above 0.97 and mean absolute error below 5.9 dB$\mu$V, showing that ML can serve as a rapid screening tool before costly measurement campaigns. For SiC converters, Li et al. \cite{Li2026} proposed a frequency-band-aware physics-informed GAN (FBA-PIGAN) for joint EMI prediction and adaptive suppression. By imposing harmonic-structure, parasitic-resonance, and high-frequency-envelope constraints, and by mapping predicted spectra to passive-filter parameters through an embedded analytical transfer function, the framework linked data-driven spectral modeling with physical filter design. On a 10 kW SiC inverter platform with 5120 measured spectra over 32 operating conditions, FBA-PIGAN achieved a mean spectral error of 2.1 dB, 93.8\% peak-frequency accuracy, and a physical-consistency score of 0.93.

AI has also been used to move mitigation from fixed post-design filtering toward adaptive, design-integrated suppression. Lu et al. \cite{Lu2026} formulated active EMI filtering for automotive electric drives as an RL problem, using a variational autoencoder for compact state representation and an exploration mechanism to tune filter parameters under changing interference profiles. The method achieved 25--30 dB attenuation improvements on experimentally measured spectra. Halidu et al. \cite{Halidu2025} used an ANN-optimized random pulse-width modulation scheme to reduce conducted EMI in three-phase voltage-source inverters while avoiding additional hardware. These studies indicate that future EMC workflows should jointly optimize layout parasitics, switching transitions, modulation, filter synthesis, and measurement uncertainty. Nevertheless, AI-based EMI prediction should be treated as an early design and margin-assessment tool; formal EMC certification still requires standardized measurement and compliance testing \cite{EMI_Mitigation_Toolset}.

\subsection{PCB Layout and Circuit Design Automation}

The physical layout of a converter is not a downstream drafting activity; it is an electrical, thermal, and EMC design variable. In high-frequency switching converters, trace routing, commutation-loop area, package parasitics, creepage and clearance, thermal vias, grounding, shielding, and component placement directly influence efficiency, switching transients, common-mode noise, heat spreading, and reliability. Expert PCB designers therefore rely on tacit knowledge accumulated from waveform debugging, post-layout extraction, EMI testing, and manufacturing feedback. AI-assisted layout automation aims to encode part of this knowledge into rule-aware generation and optimization while preserving design-rule checking and engineer review.

\begin{table*}[t!]
	\caption{Summary of AI methods for power-converter-rich electrical-system design}
	\label{tab:ai-paradigms}
	\centering
	\scriptsize
	\setlength{\tabcolsep}{3.2pt}
	\renewcommand{\arraystretch}{1.18}
	\begin{tabular}{p{2.15cm}p{3.6cm}p{3.4cm}p{4.5cm}p{2.0cm}}
		\toprule
		\textbf{AI Paradigm} & \textbf{Application Area} & \textbf{Key Strengths} & \textbf{Limitations} & \textbf{Key References} \\
		\midrule
		Supervised learning (ANNs, CNNs, KNN) & FEA surrogate modeling; EMI prediction; magnetic-component parameter estimation; efficiency and thermal mapping & High interpolation accuracy; millisecond-level inference; mature training workflows & Data-intensive; weak extrapolation; limited interpretability; requires representative simulation or measurement data & \cite{JESTIE.2022.3198504,TPEL.2020.3024914,Viarouge2024} \\
		\addlinespace[3.0pt]
		PINNs and PIML & Magnetic-field prediction; thermal analysis; lifetime modeling; multi-physics coupling & Enforces physical consistency through equation or model embedding; improves data efficiency and prediction plausibility & Requires known governing equations and boundary conditions; loss balancing can be difficult; training may be unstable & \cite{TPWRS.2022.3162473,FraunhoferPINN,APEC.2022.9773482} \\
		\addlinespace[3.0pt]
		RL and DRL & Topology discovery; sequential design decisions; active EMI filtering; simulator-feedback optimization & Explores combinatorial spaces; learns policies from interaction; supports autonomous search & Training is computationally expensive; sim-to-real transfer remains difficult; generated decisions require formal verification & \cite{ICCAD.2021.9643548,RSER.2025.116591,Lu2026} \\
		\addlinespace[3.0pt]
		GANs and generative models & EMI spectrum synthesis; PCB or layout candidate generation; inverse design & Represents high-dimensional distributions; supports design generation and inverse mapping & Training instability; mode collapse; generated candidates require physical, regulatory, and manufacturing checks & \cite{Li2026,Yang2025,EMA_EDA_PCB_AI} \\
		\addlinespace[3.0pt]
		Evolutionary and metaheuristic algorithms & Pareto optimization; magnetic geometry optimization; mixed discrete--continuous parameter search & Gradient-free global search; naturally handles nonconvex and constrained spaces & Slow convergence; high simulation cost without surrogates; sensitive to encoding and hyperparameters & \cite{Lin2025,Wang2022,Shen2024} \\
		\addlinespace[3.0pt]
		LLMs and knowledge graphs & Requirement interpretation; data-sheet extraction; SPICE-model generation; knowledge retrieval; documentation & Flexible human--machine interaction; integrates unstructured design knowledge; improves workflow accessibility & Hallucination risk; limited grounding without retrieval and simulation feedback; unsuitable for unchecked design approval & \cite{D2S_FLOW,Flux_AI_Tool,Graphwise_Statnett,Nau2025} \\
		\addlinespace[3.0pt]
		Agentic AI frameworks & Multi-tool orchestration; simulation-in-the-loop design; workflow automation from specification to report & Task decomposition; tool invocation; integration of heterogeneous models, databases, and simulators & System complexity; dependence on tool quality; traceability, verification, and liability remain unresolved & \cite{abouali2026agenticsurvey,Chen2025Collaborative,Poddar2026} \\
		\bottomrule
	\end{tabular}
\end{table*}

Generative PCB and circuit-design methods attempt to translate specifications and schematics into layout candidates that satisfy electrical and physical constraints. Yang et al. \cite{Yang2025} proposed an automatic power-electronic PCB layout framework based on generative AI, positioning it as a step toward a hardware compiler for power electronics. In industrial electronic design automation (EDA), AI-enabled platforms can generate power architectures, regulator circuits, and power trees, then optimize cost, size, and efficiency before exporting schematics for downstream workflows \cite{EMA_EDA_PCB_AI}. Reported industrial case studies further indicate that AI automation in Allegro X AI can shorten PCB design turnaround time while preserving electrical design constraints \cite{Qualcomm_Allegro_AI}. These EDA examples should be interpreted as transferable workflow evidence rather than as direct approval evidence for converter safety or EMC compliance. For power electronics, the decisive challenge is not merely producing a routable board. A credible AI layout loop must connect schematic intent, placement, routing, parasitic extraction, thermal analysis, insulation constraints, EMC screening, and manufacturability checks, and it must expose the reasons why a generated layout satisfies or violates those constraints.

\subsection{Reliability-Oriented Design Optimization}

Reliability-aware design asks whether the converter can meet its performance targets not only at commissioning, but across the mission profiles, thermal cycles, overload events, and environmental stresses expected during service. This requirement is difficult because semiconductor devices, capacitors, magnetic components, solder layers, bond wires, substrates, and cooling interfaces age under coupled electrical, thermal, and mechanical stress. Dragicevic et al. \cite{Dragicevic2019} introduced artificial-intelligence-aided automated design for reliability (ADfR), in which ML models of power loss and thermal impedance are used to estimate accumulated damage in semiconductor devices and capacitors inside the optimization loop. By making expected lifetime an explicit design objective rather than a post-design check, ADfR supports cost-, efficiency-, and reliability-constrained converter sizing.

PIML strengthens this direction by coupling neural predictors with analytical lifetime and degradation models. Such methods can support RUL estimation during early design stages, where exhaustive run-to-failure data are unavailable \cite{APEC.2022.9773482}. Embedded physical constraints reduce the risk of unrealistic degradation trajectories, which is particularly important for SiC MOSFETs and GaN HEMTs because long-term field data remain more limited than for mature silicon IGBT technologies. Multi-objective ADfR can also evaluate initial hardware cost, conversion efficiency, thermal margin, and end-of-life reliability in one optimization framework. The unresolved difficulty is interaction modeling: bond-wire liftoff, solder fatigue, gate-oxide degradation, capacitor aging, thermal-interface deterioration, and mission-profile variability are not independent mechanisms. AI can help identify and approximate these couplings, but the resulting lifetime predictions require conservative uncertainty margins and validation against accelerated-aging and field data before they can support safety-critical design decisions.

\subsection{Knowledge-Driven and Agentic Design Assistance}

Component selection and knowledge management are often underestimated bottlenecks in converter design. Engineers must interpret data sheets, application notes, design guides, safety standards, simulation records, and previous test reports to select semiconductors, magnetic cores, capacitors, gate drivers, sensors, and cooling components that jointly satisfy electrical, thermal, mechanical, reliability, and supply-chain constraints. Natural-language processing, knowledge graphs, retrieval-augmented generation (RAG), and LLMs are being investigated to reduce this manual burden and to make design knowledge more searchable, traceable, and reusable.

D2S-FLOW uses LLMs to extract electrical parameters from device data sheets and generate SPICE models with high precision and efficiency \cite{D2S_FLOW}. Its attention-guided document focusing, hierarchical document-enhanced retrieval, and heterogeneous named-entity normalization mechanisms are designed to handle unstructured documents, multi-page tables, and inconsistent naming conventions. Reported results include an exact-match score of 0.86, an F1 score of 0.92, and a 38\% reduction in API-token consumption relative to conventional RAG systems. LLM-based design assistants can also support natural-language queries such as ``find a GaN HEMT with $R_{DS(on)}$ below 50 m$\Omega$,'' while knowledge-graph platforms provide semantic access to complex design models and time-series databases \cite{Flux_AI_Tool,Graphwise_Statnett}. In this role, the model is less a numerical predictor than an interface between human intent, unstructured documents, component databases, and executable design tools.

Recent LLM and agentic AI systems extend knowledge assistance toward workflow orchestration. The power electronics GPT (PE-GPT) paradigm proposed by Lin et al. \cite{Lin2025} combines a multimodal LLM, a specialized power-electronics knowledge base, RAG, metaheuristic algorithms, a model zoo, and a simulation repository. In dual-active-bridge modulation and buck-converter parameter-design case studies, PE-GPT improved correctness by 22.2\% relative to human experts and by 35.6\% relative to other leading LLMs, while improving consistency and reducing hallucination. Chen et al. \cite{Chen2025Collaborative} proposed a collaborative DRL--LLM framework in which the LLM interprets natural-language requirements and initializes the design state, while the DRL agent refines topology generation and layout routing through reward-based optimization.

The decisive requirement for such systems is simulation feedback. Without feedback, an LLM or agent can produce plausible but invalid circuits; with feedback, it can revise candidates against executable evidence. Nau et al. \cite{Nau2025} evaluated LLM-based switched-mode power-supply design on 269 manually created benchmark tasks and found that combining SPICE feedback with LLM reasoning increased the solve rate from 15\% to 91\%. Parameter tuning was largely solvable, whereas topology adaptation remained more difficult, indicating that current systems are stronger at constrained refinement than at open-ended invention. Agentic frameworks generalize this idea by coordinating geometric parameterization, FEA surrogates, optimizers, circuit simulators, and documentation tools under a planning model \cite{wu2024autogen}. The HeaRT framework \cite{Poddar2026}, although developed for analog and mixed-signal optimization, illustrates the value of hierarchical circuit reasoning, reporting F1-score improvements of 13.5\% for subcircuits and 37.8\% for loops over few-shot prompting baselines, as well as more than threefold faster convergence under specification shifts.

Table \ref{tab:ai-paradigms} compares the main AI paradigms used in converter design. A consistent pattern emerges across the table: AI is becoming most valuable when it is coupled to constraints, uncertainty estimates, and executable verification. The credible near-term trajectory is therefore hybrid autonomy, not unsupervised design approval.

Overall, AI-enabled converter design is progressing from isolated surrogate models toward integrated and verifiable design environments. The main unresolved issues are extrapolation beyond training distributions, uncertainty quantification, reproducible benchmarking, manufacturability of generated layouts, safety and EMC margins, and traceable validation of AI-generated topologies or component choices. Progress will depend less on any single AI paradigm than on principled coupling among physics-informed learning, optimization, simulation feedback, experimental datasets, standards-aware design rules, and human review. This coupling is essential if AI is to mature from a rapid exploration tool into a trustworthy design infrastructure for power-converter-rich electrical systems.

\section{Artificial Intelligence for Control}

AI methods are applied to control tasks in power converters and power systems to approximate plant dynamics, emulate control laws, tune controller parameters, and reduce online optimization burden. These methods are mainly studied for systems with nonlinear dynamics, fast sampling requirements, and operating uncertainty. The reviewed literature is grouped according to the position of the learning module in the control loop and the kind of deployment evidence it can provide. Supervised learning usually serves as an offline surrogate of models or controllers. Reinforcement learning learns a control policy through interaction with a converter or grid environment. Safety-aware learning introduces voltage, current, frequency, and power-flow limits during training or online execution. Representative studies are listed in Table \ref{tab_ai_control_studies}; the table is intended as a role-based map rather than a direct benchmark because the cited works use different plants, sampling rates, constraints, and validation platforms.

\begin{table*}[t!]
	\caption{Summary of AI methods for power-converter-rich electrical-system control}
	\label{tab_ai_control_studies}
	\centering
	\scriptsize
	\setlength{\tabcolsep}{3.5pt}
	\renewcommand{\arraystretch}{1.22}
	\begin{tabular}{p{2.65cm}p{3.80cm}p{3.65cm}p{5.65cm}}
		\toprule
		\textbf{Research Focus} & \textbf{Representative Method} & \textbf{Main Application} & \textbf{Control Role and References} \\
		\midrule
		\multicolumn{4}{c}{Supervised learning and imitation learning} \\
		\midrule
		Neural approximation for converter control & Deep machine learning control; deep-learning-based MPC; steady-state neural modeling & dc/dc converters; resonant converters; CLLC converters in dc distribution systems & Converter dynamics or control laws are learned from data to support real-time regulation and predictive control under nonlinear operating conditions \cite{TPEL.2020.2977765,TII.2020.2969729,APEC43599.2022.9773436}. \\
		\addlinespace[3.5pt]
		Neural surrogates for predictive control & Long horizon MPC; machine learning emulation of MPC; ANN assisted MPC & PMSM drives; modular multilevel converters; general power converters & Predictive control actions are approximated by neural models for long horizon prediction, voltage regulation, submodule balancing, and switching selection \cite{TPEL.2022.3172681,TIE.2020.3038064,TIE.2021.3076721}. \\
		\addlinespace[3.5pt]
		Learning assisted finite set and adaptive control & Model free adaptive predictive control; finite control set learning predictive control; iterative learning predictive control & Power converters with uncertainty or repetitive operation & Local dynamics, switching decisions, and repetitive tracking errors are used to update predictive control actions with fewer online searches \cite{TIE.2022.3208594,TIE.2023.3303646,TPEL.2022.3194518}. \\
		\addlinespace[3.5pt]
		Imitation learning for fast switching control & Physics informed imitation learning; supervised imitation learning of finite set MPC & Power converters and power electronics systems & Expert control actions or finite set MPC decision boundaries are reproduced, with circuit laws or topology constraints included when available \cite{JESTPE.2026.3673944,TIE.2020.2969116}. \\
		\addlinespace[2.0pt]
		\midrule
		\multicolumn{4}{c}{Reinforcement learning control} \\
		\midrule
		Autonomous and transferable RL control & Deep reinforcement learning; transfer reinforcement learning; domain adaptation based transfer learning & Power grid voltage control; dc/dc converters; dc microgrids & Policies are learned or transferred for voltage control, constant power load stabilization, and operation under new converter or microgrid conditions \cite{TPWRS.2019.2941134,JESTPE.2022.3189078,TIE.2022.3192676,TIE.2025.3572981}. \\
		\addlinespace[3.5pt]
		RL assisted tuning of predictive and adaptive control & Adaptive horizon seeking; variable self tuning horizon control; parameter self configuration; DRL based adaptive control & dc/dc converters; autonomous dc microgrids; nonlinear systems and PMSM drives & Reinforcement learning is used to tune prediction horizons, nonsmooth control parameters, or adaptive controller parameters under changing loads and disturbances \cite{TCSI.2023.3325590,JESTPE.2022.3225264,JESTPE.2023.3347515,TII.2024.3507191}. \\
		\addlinespace[2.0pt]
		\midrule
		\multicolumn{4}{c}{Safety constrained and robust learning control} \\
		\midrule
		Projection and shielding based safe control & Safe multi agent DRL; DNN assisted projection; physics shielded multi agent DRL & Inverter based renewable energy sources; distribution grids with PV and battery storage & Candidate actions are projected, accelerated, or replaced so that voltage and network operating limits are considered during active voltage control \cite{TSTE.2023.3341632,TPWRS.2023.3336614,TSG.2022.3228636}. \\
		\addlinespace[3.5pt]
		Barrier functions, constrained learning, and safety verification & Control barrier functions; explainable safety-aware DRL; constrained RL; distributed safety verification & Multi-area frequency control; stochastic dynamic optimal power flow; networked control systems & Safety boundaries, power-flow constraints, invariant sets, and tube-based tracking are introduced to support safe learning and verification \cite{TPWRS.2024.3483994,TASE.2025.3554431,TPWRS.2023.3326121,TCNS.2021.3074218}. \\
		\addlinespace[3.5pt]
		Safe converter and inverter control & Safety-enhanced self-learning; safe reinforcement learning; safe data-driven control framework; AI-assisted adaptive PI control & Power converters; grid-forming inverters; smart grids; wireless power transfer systems & Electrical limits, Lyapunov-related information, common evaluation settings, or adaptive gain tuning are included in learning-based control \cite{TIE.2024.3363759,MPCE.2023.000882,TSG.2025.3616402,TIA.2025.3572075}. \\
		\bottomrule
	\end{tabular}
\end{table*}

For converter control, the decisive question is not whether a learning algorithm can improve a simulation metric in isolation. The more important question is whether the learned component can be placed in a closed-loop architecture with known time scale, constraint handling, stability or safety evidence, and implementation cost. The subsections below therefore distinguish learning used as a controller surrogate, learning used to synthesize policies through interaction, and learning constrained by projection, shielding, barrier functions, or verification layers.

\subsection{Supervised Learning Methods for Control}

\begin{figure}[htbp!]
\centerline{\includegraphics[width=1.0\linewidth]{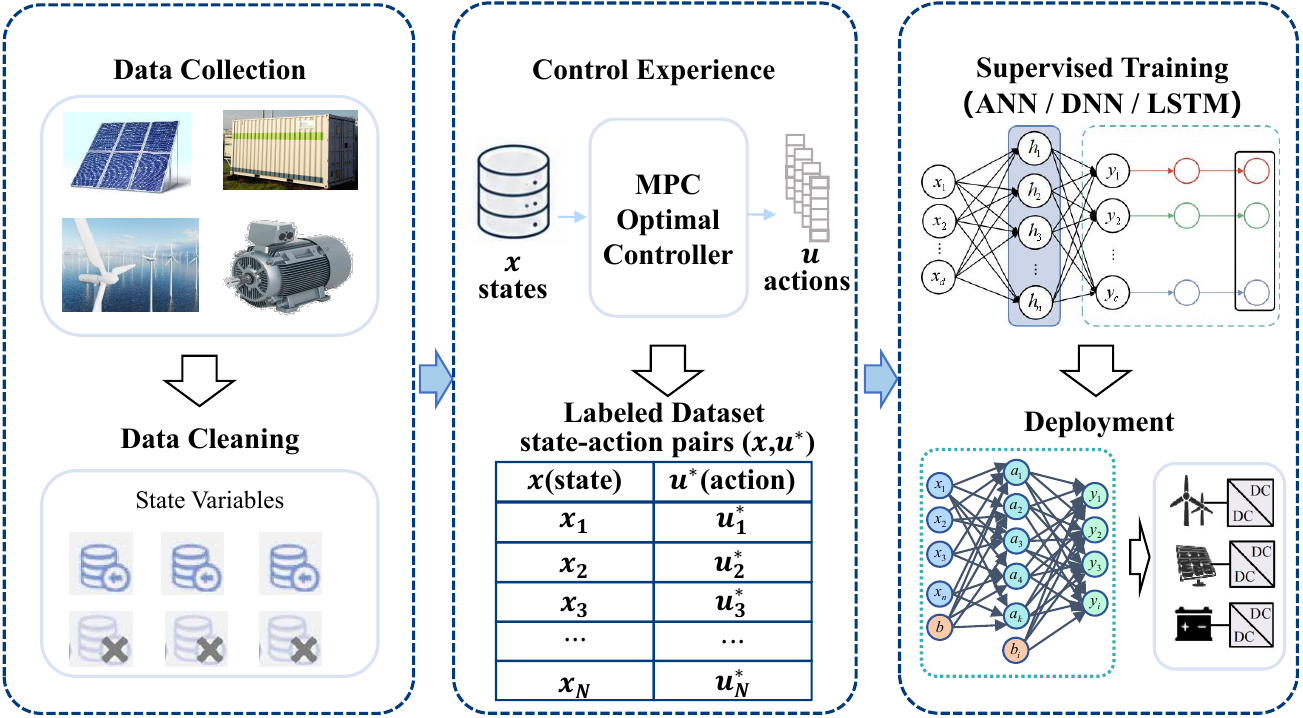}}
\caption{Workflow of supervised-learning-based control.}
\label{fig_supervised_learning_control}
\end{figure}

Fig. \ref{fig_supervised_learning_control} presents a typical supervised learning workflow for converter and power system control. Operational data are first collected from simulations, analytical models, or experiments, and the measured variables are processed as candidate states. Control experience is then generated by an MPC scheme or another optimal controller, which provides labeled state action pairs for training. The trained ANN, DNN, or LSTM model is deployed as a fast mapping from measured states to control commands. In such a workflow, the learned model is mainly used to reproduce a known control law or a system response within the region covered by the training data.

Supervised learning is therefore commonly used when control trajectories, switching states, or converter responses are available before deployment. In dc/dc converter control, deep machine learning is used for real-time regulation under nonlinear operating conditions \cite{TPEL.2020.2977765}. For resonant converters, deep-learning-based MPC is used to approximate the predictive control law and to reduce online computation while tracking behavior is retained \cite{TII.2020.2969729}. Long-horizon MPC for PMSM drives is assisted by deep neural networks so that a longer prediction window can be considered within the sampling period \cite{TPEL.2022.3172681}. Similar concepts are used for CLLC dc/dc resonant converters in dc distribution systems, where deep-learning-based steady-state modeling is coupled with MPC under load variation \cite{APEC43599.2022.9773436}.

Neural surrogates are also used to emulate computationally demanding predictive controllers. For modular multilevel converters, machine learning emulates MPC actions for submodule balancing and voltage regulation, which reduces execution time in the control loop \cite{TIE.2020.3038064}. Artificial neural networks are embedded in MPC structures for power converters so that switching decisions can be obtained with a lower online burden \cite{TIE.2021.3076721}. Predictor based model free adaptive predictive control estimates local converter dynamics online and supports control when internal parameters are uncertain \cite{TIE.2022.3208594}. Finite control set learning predictive control combines learning with discrete switching state selection to form control actions with fewer repeated searches over the switching set \cite{TIE.2023.3303646}. Iterative learning predictive control uses error information from repeated switching cycles to refine control actions in repetitive converter operation \cite{TPEL.2022.3194518}. Physics informed imitation learning includes circuit laws and topology constraints in the training process for adaptive power converter control \cite{JESTPE.2026.3673944}. Supervised imitation learning further reproduces the decision boundaries of finite set MPC and supports fast switching selection on standard digital control hardware \cite{TIE.2020.2969116}.

\subsection{Reinforcement Learning Based Autonomous Control}

\begin{figure}[!t]
\centerline{\includegraphics[width=0.6\linewidth]{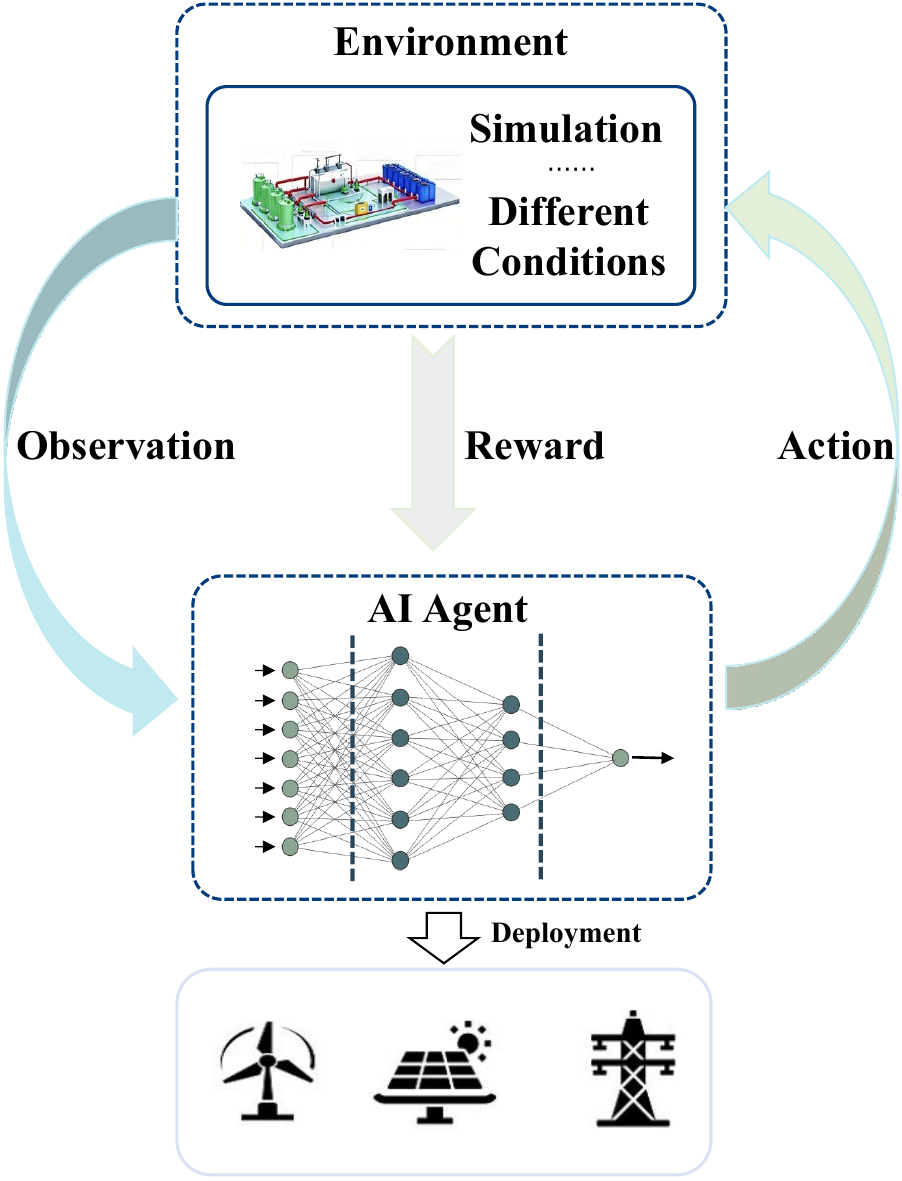}}
\caption{Workflow of reinforcement-learning-based control.}
\label{fig_reinforcement_learning_control}
\end{figure}

Fig. \ref{fig_reinforcement_learning_control} shows the closed loop interaction between an AI agent and a converter or grid environment. The environment may be represented by a simulation model with different operating conditions, and it returns observations and rewards to the agent. The agent sends control actions back to the environment, and the policy is updated according to the observed control performance. After training, the learned policy is deployed for renewable energy sources, converters, or grid equipment. Labeled optimal actions are not required during policy training, while reward design and environment fidelity can affect the resulting controller.

Reinforcement learning is used when the controller is represented as a policy that maps observed system states to control actions. The formulation can be useful for nonlinear systems and operating points with uncertain parameters, although training cost and experimental validation remain important concerns. In power grid operation, deep reinforcement learning is applied to autonomous voltage control under topological changes and load fluctuations \cite{TPWRS.2019.2941134}. For dc/dc boost converters feeding constant power loads, reinforcement learning is used for large signal stabilization when negative impedance characteristics challenge linear controllers \cite{JESTPE.2022.3189078}. Transfer reinforcement learning with duty ratio mapping is applied to buck converter regulation so that a trained policy can be adapted to new operating conditions \cite{TIE.2022.3192676}. In dc microgrids, domain adaptation based transfer learning supports the implementation of reinforcement learning controllers across different system configurations \cite{TIE.2025.3572981}.

Reinforcement learning is also used to tune predictive control structures. Deep reinforcement learning determines the prediction horizon for generalized predictive control in dc/dc converter applications \cite{TCSI.2023.3325590}. A variable self tuning horizon mechanism further adjusts the horizon of generalized dynamic predictive control for boost converters with constant power loads \cite{JESTPE.2022.3225264}. Parameter self configuration based on deep reinforcement learning is used for nonsmooth control of autonomous dc microgrids without repeated manual tuning \cite{JESTPE.2023.3347515}. For a class of nonlinear systems with mismatched disturbances, deep reinforcement learning is used to adapt controller parameters and is further evaluated in drive related applications \cite{TII.2024.3507191}.

\subsection{Safety-Aware and Robust AI Control}

\begin{figure}[!t]
\centerline{\includegraphics[width=1.0\linewidth]{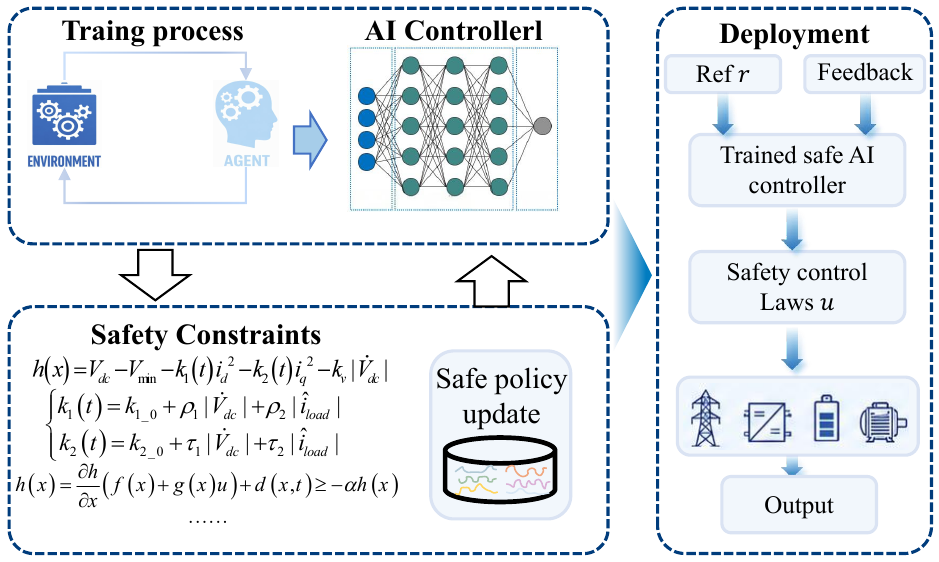}}
\caption{Workflow of safety-constrained learning-based control.}
\label{fig_safety_constrained_learning_control}
\end{figure}

Fig. \ref{fig_safety_constrained_learning_control} summarizes a safety constrained learning architecture for physical control systems. During training, the AI controller is updated together with safety constraints derived from electrical limits, barrier functions, or other control laws. Candidate actions are modified, filtered, or penalized when voltage, current, frequency, or state constraints may be violated. During deployment, reference commands and feedback are processed by the trained safe controller before control laws generate the output command. The diagram highlights a common feature of safe learning methods, where safety information is included both in policy training and in online control execution.

Safety-aware AI control receives increasing attention because learned actions are applied to physical devices with limited current, voltage, frequency, and thermal margins. Across these studies, constraints are introduced through projection, shielding, barrier functions, constrained optimization, or safe learning in predictive control. Projection-based deep reinforcement learning constrains inverter control actions so that voltage limits and distribution-grid operating ranges are respected \cite{TSTE.2023.3341632}. DNN-assisted projection reduces the computational burden of the projection step in distribution-grid control \cite{TPWRS.2023.3336614}. Physics-shielded multi-agent reinforcement learning modifies active voltage-control actions by replacing unsafe actions with shielded alternatives in photovoltaic and battery energy storage systems \cite{TSG.2022.3228636}.

Barrier-function-based approaches provide another way to include safety information in learning-based control. Control barrier functions are used to define time-varying safety boundaries for multi-area frequency control without relying on an accurate nominal model \cite{TPWRS.2024.3483994}. Explainable and safety-aware reinforcement learning further uses neural-network gradient information to interpret control actions in nonlinear discrete-time systems \cite{TASE.2025.3554431}. Constrained reinforcement learning directly learns policies for stochastic dynamic optimal power flow while generation, storage, and power-flow constraints are considered \cite{TPWRS.2023.3326121}. For networked control systems, plug-and-play distributed safety verification uses robust invariant sets and tube-based tracking to handle bounded uncertainty and changing interconnections \cite{TCNS.2021.3074218}.

Safety constrained learning is also studied in converter and inverter applications. Safety enhanced self learning includes electrical limits in the policy update process for optimal power converter control \cite{TIE.2024.3363759}. Safe reinforcement learning is applied to grid forming inverter frequency regulation with Lyapunov related stability information \cite{MPCE.2023.000882}. Common evaluation frameworks are developed for safe data driven smart grid control so that learning controllers can be compared under shared operating scenarios \cite{TSG.2025.3616402}. In wireless power transfer systems, artificial intelligence assisted adaptive PI control is used to tune controller gains under load and coupling coefficient variations \cite{TIA.2025.3572075}.

Taken together, these studies suggest that learning-based control becomes most credible when the learning component is bounded by physical constraints and by a conventional control or verification layer. For an IEEE power-electronics audience, future reviews and benchmarks should therefore report not only reward or tracking error, but also sampling period, computation platform, constraint-violation rate, training domain, disturbance range, and whether validation is simulation-only, HIL-based, or experimental.

\section{Artificial Intelligence for Operations}

At the operation layer, the connection to power-converter-rich systems must be made explicit. AI is not reviewed here as a generic smart-grid tool, but as a means of coordinating inverter-interfaced DERs, storage converters, flexible loads, and distribution networks whose behavior is shaped by fast voltage dynamics, low inertia, bidirectional power flow, communication constraints, and cyber-physical exposure. The operational value of AI therefore lies in converting heterogeneous measurements into dispatch, forecasting, observability, voltage-control, maintenance, and security decisions while respecting the limits of converter-interfaced assets.

\subsection{Microgrid Operations}

In microgrids (MGs), ML, DL, and RL are mainly used to manage renewable intermittency, DER proliferation, storage operation, and converter-dominated power exchange. The reviewed applications include optimal dispatch, forecasting, resilience enhancement, predictive maintenance, and supporting digital technologies.
\vspace{0.3em}
\subsubsection{Optimal Dispatch and Energy Management} 
\indent The energy management system (EMS) is central to microgrid operation, coordinating DERs, storage, and loads to minimize cost while ensuring reliability \cite{cordova2023aggregate}. Conventional optimization methods often struggle with non-convex optimal power flow and forecast uncertainty, limiting real-time applicability in dynamic environments \cite{wang2024nonconvex}.

RL and deep reinforcement learning (DRL) address these challenges by learning optimal control policies directly from system interactions. Algorithms such as deep deterministic policy gradient and twin delayed deep deterministic policy gradient (TD3) handle continuous control variables (e.g., inverter setpoints) \cite{dominguez2023twin}, while deep Q-network is effective for discrete decisions like unit commitment, enabling real-time coordination of photovoltaic, battery systems, and grid exchange \cite{asghar2025deep}.

\begin{figure}[!t]
\centering
\includegraphics[width=\columnwidth]{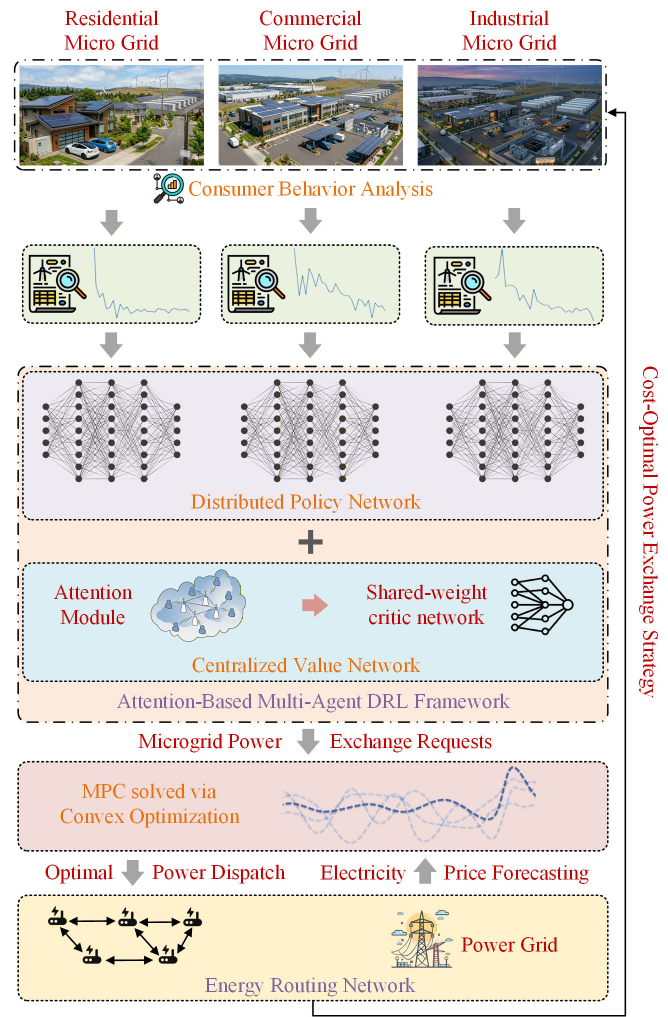}
\caption{Attention-based multi-agent DRL framework for microgrid energy management and power exchange \cite{zhang2023multi}.}
\label{fig:multiagent-drl}
\vspace{-1.6em}
\end{figure}

Furthermore, multi-agent reinforcement learning enables decentralized coordination among interconnected microgrids, improving reliability and scalability \cite{li2023multi}, as shown in Fig. \ref{fig:multiagent-drl}. Hybrid approaches combining artificial neural networks with metaheuristics such as particle swarm optimization (PSO) and genetic algorithms enhance convergence and solution quality. An ANN-PSO model, for instance, has achieved low normalized mean square error, while fuzzy logic supports interpretable decision-making under uncertainty \cite{talaat2023artificial}. Together, these approaches address typical EMS objectives such as cost, degradation, and emissions while ensuring compliance with power balance and operational constraints.

\vspace{0.3em}
\subsubsection{Operational Forecasting} 
\indent Accurate forecasting of load demand, renewable generation, and electricity prices is essential for proactive microgrid management \cite{r2024machine}. However, variability in weather conditions and data uncertainty make reliable prediction a challenging task.

Deep learning models such as long short-term memory (LSTM) networks and convolutional neural networks (CNNs) are widely used due to their ability to capture temporal and spatial dependencies \cite{ozcanli2022islanding}. Reported LSTM-based models show strong forecasting performance when historical and meteorological inputs are available, but the achieved accuracy depends on the local data set, prediction horizon, and renewable-generation profile \cite{li2023deep}, \cite{wang2024live}.

For lower computational complexity, models such as extreme gradient boosting (XGBoost) provide robust alternatives when only limited features are available \cite{wang2024live}. Probabilistic approaches such as generative adversarial networks (GANs) generate scenarios for stochastic optimization \cite{mansour2024wasserstein}, while hybrid wavelet-neural network models improve forecasting under highly variable conditions \cite{aly2022hybrid}. Despite these advances, challenges persist under sudden weather shifts or incomplete data, motivating continued research into adaptive forecasting architectures.
\vspace{0.3em}
\subsubsection{Resilience Enhancement During Extreme Events} 
\indent Resilience is a critical requirement for microgrids operating under extreme conditions, where rapid response and adaptability are essential. Traditional optimization-based restoration methods often suffer from high computational time, limiting their effectiveness in real-time scenarios.

DRL enables fast and adaptive decision-making for actions such as islanding and load restoration, and reported studies show sub-second response in simulation-based restoration settings \cite{jeyaraj2023deep}.

Multi-agent reinforcement learning (MARL) further enhances resilience by coordinating distributed resources such as mobile energy storage systems (MESSs), electric vehicles (EVs), and repair crews (RCs) while considering electrical and logistical constraints \cite{wang2023towards}. Reported MARL approaches can approach centralized restoration performance in test systems \cite{ahsan2025multi}, although the simulation-to-reality (sim-to-real) gap still requires robust training and transfer learning strategies.
\vspace{0.3em}
\subsubsection{Predictive Maintenance and Power Quality Management} 
\indent The reliability of microgrids depends on the health of critical components such as batteries, converters, and transformers, which are subject to thermal and electrical stresses. Traditional maintenance strategies are often reactive and inefficient.

AI-based predictive maintenance enables condition-based monitoring by analyzing operational data to detect early degradation and estimate remaining useful life (RUL) \cite{sayal2024ai}. Machine learning models such as artificial neural networks (ANNs), support vector machines (SVMs), and decision trees process parameters like voltage, current, temperature, and state-of-charge, while LSTM models effectively capture temporal degradation trends \cite{omo2026gmdh}, \cite{mohamed2022dynamic}.

In parallel, AI-driven power quality (PQ) management identifies disturbances such as voltage sags, swells, and harmonics \cite{zahid2026ai}. Signal processing techniques, including wavelet transform and fourier transform, combined with classifiers like CNNs and SVMs enable fast detection, although robustness under noisy conditions \cite{cano2024integrating} and limited labeled data remains a key challenge.

\subsubsection{Integration with Enabling Digital Technologies}
\begin{figure}[!t]
\centering
\includegraphics[width=\columnwidth]{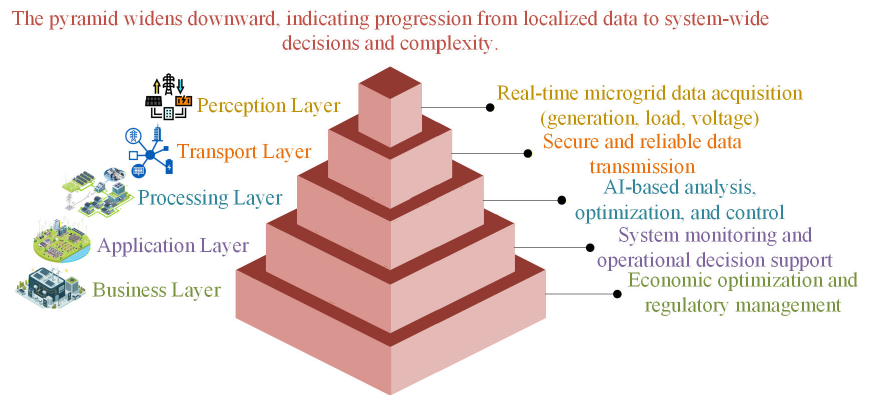}
\caption{Layered IoT-AI architecture for microgrid data, communication, and optimization \cite{hadi2025artificial}.}
\label{fig:IoTai}
\end{figure}

\indent The effectiveness of AI in microgrid operations is significantly enhanced through integration with digital technologies that provide data, connectivity, and computational support. The Internet of Things (IoT) enables real-time sensing, communication, and monitoring across multiple system layers \cite{sitharthan2023smart}, as presented in Fig. \ref{fig:IoTai}.

Federated learning (FL) facilitates collaborative model training across distributed entities without sharing raw data, preserving privacy while introducing communication overhead \cite{veerasamy2023blockchain}. Blockchain technology ensures secure and transparent energy transactions through decentralized and tamper-resistant ledgers, particularly in peer-to-peer (P2P) energy systems \cite{veerasamy2024blockchain}.

Additionally, digital twins (DTs), which are virtual replicas of physical microgrids, can be combined with AI to support real-time simulation, optimization, and predictive analysis \cite{fan2023energy}. These technologies collectively enhance system efficiency, resilience, and autonomy, supporting the evolution of intelligent and decentralized energy systems.

\subsection{Power System Operations}
Building on advancements in microgrid operations, AI is also being investigated at broader power-system scales where renewable penetration, converter-interfaced resources, and real-time operational demands increase complexity. Conventional methods such as optimal power flow and load frequency control remain essential, but they may face computational or modeling difficulties under strong nonlinearity, uncertainty, and frequent operating changes. AI, especially DRL, is therefore studied as a complementary decision layer for adaptive optimization, stability support, and resilience analysis.

\subsubsection{Real-Time Optimal Power Flow and Economic Dispatch} %
\indent Real-time optimal power flow is essential for responding to rapidly changing system conditions; however, conventional interior-point methods are computationally intensive and unsuitable for sub-minute dispatch due to iterative convergence and sensitivity to initialization \cite{pervez2025computationally}.

DRL addresses this by learning a direct mapping from system states (e.g., bus voltages, loads, generation levels) to control actions (generator outputs, voltage setpoints). Lagrangian-based DRL incorporates power balance and network constraints into the reward via penalty terms, and reported studies show near-optimal generation costs with substantially reduced computation time relative to iterative solvers \cite{yan2020real}.

Beyond numerical objectives, AI also enables early exploration of qualitative operational constraints. Integrating DRL with generative pre-trained transformer (GPT)-based agents allows grid-code information, such as voltage-security and contingency limits, to be represented in the decision process, although such language-conditioned operation remains an emerging research direction rather than a certified dispatch procedure \cite{yan2023real}.

\begin{figure}[!t]
\centering
\includegraphics[width=\columnwidth]{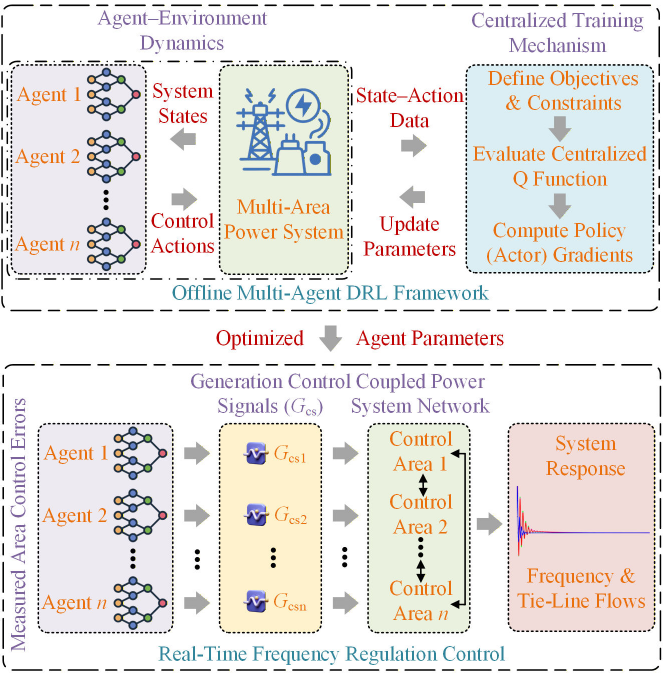}
\caption{MADRL-based framework for load frequency control in multi-area power systems \cite{9108633}.}
\label{fig:madrl}
\end{figure}

\subsubsection{Multi-Area Load Frequency Control and Stability} 
\indent Load frequency control in multi-area power systems requires coordinated regulation of system frequency and tie-line power exchanges under load disturbances. Conventional PID-based controllers lack coordination and struggle with nonlinearities such as turbine dynamics and time delays \cite{davoudkhani2024robust}.

Multi-agent deep reinforcement learning (MADRL), particularly multi-agent deep deterministic policy gradient, enables cooperative control using centralized training and decentralized execution \cite{9108633}, as illustrated in Fig.~\ref{fig:madrl}. Agents rely on local area control error (ACE) while considering nonlinear constraints such as generation rate constraints and governor dead bands.

Simulation studies on benchmark systems (e.g., IEEE multi-area networks) show that MADRL reduces mean absolute ACE by $\sim$31\% compared to tuned PID controllers and improves disturbance rejection and settling time, demonstrating superior dynamic performance and scalability \cite{wang2024live}.

\subsubsection{Integration of Variable Renewable Energy Sources} 
\indent The integration of variable renewable energy (VRE) into power systems introduces balancing, profile, and grid-related costs due to forecast errors, temporal variability, and location-dependent constraints, which can reach 15-41 EUR/MWh at moderate penetration levels \cite{hua2022applications}.

AI mitigates these challenges through enhanced forecasting and optimization. Machine learning models improve solar and wind forecasting accuracy by up to 30\%, while advanced demand forecasting achieves errors as low as 2.28\%, reducing reserve requirements and improving dispatch scheduling \cite{hua2022applications}.

Additionally, AI-driven optimization supports demand response, storage control, and asset management. Applications such as DRL-based energy optimization and machine learning-based battery degradation modeling ($\sim$9\% prediction error) improve lifecycle efficiency, making VRE integration more economically sustainable \cite{hua2022applications}.

\subsubsection{Security, Vulnerability, and Resilience of AI-Driven Operations} 
\indent Although DRL offers significant operational benefits, its deployment in power systems introduces new security vulnerabilities. Deep neural networks, which form the core of DRL agents, are sensitive to adversarial perturbations, where small, carefully crafted changes in input states (e.g., load, voltage, or topology data) can lead to incorrect control actions and potential system instability \cite{ren2024transferable}.

Studies indicate that perturbations of around 10\% in system state variables can degrade DRL performance by over 30\% and increase failure rates under targeted attacks, particularly in topology optimization and dispatch applications \cite{boza2021artificial}. This highlights the susceptibility of DRL-based control to structured and time-coordinated disturbances.

To mitigate these risks, vulnerability metrics such as the action preference value $p(s_t)$ representing the gap between the highest and second-highest action probabilities are used to identify critical operating conditions \cite{boza2021artificial}. Such indicators enable early detection of instability and support the development of more robust DRL models through improved training and secure deployment strategies.

\subsection{Distribution System Operations}
Following system-level AI integration, distribution systems are rapidly transforming with widespread integration of DERs, smart meters, and advanced sensors. These systems are especially relevant to a converter-rich review because inverter-based DERs create reverse power flow, local voltage excursions, topology-dependent control interactions, and privacy-sensitive data streams. AI, especially ML and DL, can improve observability, manage variability, and support fast decentralized decision-making when the available measurements are incomplete or heterogeneous.

\subsubsection{Distribution System State Estimation} 
\indent Distribution system state estimation (DSSE) is essential for real-time monitoring and control but is challenged by limited measurements, high network complexity, and data quality issues \cite{ahmad2018distribution}. Traditional weighted least squares (WLS) methods struggle with missing data, extreme values, and under-determined conditions, reducing estimation reliability \cite{khashei2025mean}.

\begin{table*}[!t]
	\caption{Summary of AI Methods for Power-Converter-Rich Electrical System Operations}
	\label{tab:ai_power_systems}
	\centering
	\scriptsize
	\setlength{\tabcolsep}{3.2pt}
	\renewcommand{\arraystretch}{1.18}
	\begin{tabular}{p{2.35cm}p{2.35cm}p{5.10cm}p{4.10cm}p{2.05cm}}
		\toprule
		\textbf{Application Area} & \textbf{AI Technique} & \textbf{Methodology} & \textbf{Key Outcome} & \textbf{Key References} \\
		\midrule
		\multicolumn{5}{c}{\textbf{Microgrid Operations}} \\
		\midrule
		Optimal dispatch and EMS & DRL (DDPG, SAC, TD3) & DRL-based real-time DER dispatch and ESS scheduling under renewable uncertainty & Reduces operating cost and battery degradation while maintaining power balance & \cite{li2023deep,hadi2025artificial} \\
		\addlinespace[3.0pt]
		Multi-microgrid coordination & MADDPG, MATD3 & Multi-agent DRL with centralized training and decentralized coordination & Enhances resilience and minimizes operating cost with near-optimal performance & \cite{qiu2024artificial} \\
		\addlinespace[3.0pt]
		Hybrid AI optimization & ANN with PSO/GA & ANN forecasting integrated with metaheuristic optimization for EMS scheduling & Fast convergence with low scheduling error (NMSE $\approx$ 1.10\%) & \cite{talaat2023artificial} \\
		\addlinespace[3.0pt]
		Load and PV forecasting & LSTM, CNN, XGBoost & Deep learning models for renewable generation and load forecasting & High forecasting accuracy with $R^2$ up to 0.9831 & \cite{wang2024live,li2023deep} \\
		\addlinespace[3.0pt]
		Resilience enhancement & Offline-trained DRL & DRL-based adaptive islanding and restoration strategy & Near-optimal restoration with sub-millisecond response & \cite{qiu2024artificial} \\
		\addlinespace[3.0pt]
		Predictive maintenance & LSTM, SVM, DT & Data-driven remaining useful life prediction using operational data & Reduces unexpected outages and maintenance cost & \cite{hadi2025artificial,talaat2023artificial} \\
		\addlinespace[3.0pt]
		Power quality diagnosis & Wavelet with SVM/ANN/CNN & Wavelet-based feature extraction with AI-driven PQ disturbance classification & Classification accuracy up to 99.93\% & \cite{hadi2025artificial} \\
		\midrule
		\multicolumn{5}{c}{\textbf{Power System Operations}} \\
		\midrule
		Real-time OPF & DRL (DDPG) & Constraint-aware DRL framework for real-time AC-OPF & Near-optimal solution with $\approx$99.8\% lower computation time & \cite{yan2020real} \\
		\addlinespace[3.0pt]
		OPF with linguistic constraints & DRL with GPT agent & LLM-assisted interpretation of grid-code constraints & Improves operational compliance while preserving feasibility & \cite{yan2023real} \\
		\addlinespace[3.0pt]
		Multi-area load frequency control & MADDPG & Cooperative DRL using local ACE signals for distributed frequency control & ACE reduction of $\approx$31\% compared with PID control & \cite{yan2020multi} \\
		\addlinespace[3.0pt]
		VRE forecasting and balancing & SVR, ANN, XGBoost & ML-based probabilistic forecasting for wind, solar, and demand & Reduces balancing cost and improves renewable utilization & \cite{boza2021artificial,rocha2021artificial} \\
		\addlinespace[3.0pt]
		VRE asset optimization & DRL, ML & RL and ML for cooling optimization and battery degradation prediction & Cooling energy reduction up to 40\% & \cite{boza2021artificial} \\
		\addlinespace[3.0pt]
		Grid cost optimization & RL, predictive ML & RL-based feeder reconfiguration and predictive maintenance & Operational expenditure reduction up to 60\% & \cite{boza2021artificial} \\
		\addlinespace[3.0pt]
		Vulnerability assessment & FGSM-based adversarial ML & Adversarial perturbation analysis for DRL controller security & Identifies cyber vulnerabilities and reliability degradation & \cite{zheng2021vulnerability} \\
		\midrule
		\multicolumn{5}{c}{\textbf{Distribution System Operations}} \\
		\midrule
		Distribution state estimation & Noise-trained GAN & GAN-based DSSE robust against noisy and missing measurements & Outperforms conventional WLS under corrupted data & \cite{liu2025noise} \\
		\addlinespace[3.0pt]
		Topology identification & Correlation, DT, MLP, ADMM & Data-driven feeder topology and phase connectivity identification & Improves topology accuracy and network observability & \cite{barja2021artificial} \\
		\addlinespace[3.0pt]
		Volt-VAR control & MLG-MATD3 & GCN-enabled multi-agent DRL for coordinated Volt-VAR optimization & Voltage violations maintained below 1\% & \cite{ge2025autonomous} \\
		\addlinespace[3.0pt]
		Decentralized voltage control & MADDPG, MAPPO, IPPO & Distributed multi-agent DRL for reactive power and voltage coordination & Enhances scalability and voltage regulation performance & \cite{barja2021artificial} \\
		\addlinespace[3.0pt]
		Privacy-preserving TSO--DSO coordination & NN with quadratic regression & NN-based feasible region representation for secure TSO--DSO OPF coordination & Achieves $\approx$0.36\% cost deviation with full feasibility & \cite{dindar2026privacy} \\
		\addlinespace[3.0pt]
		ML-accelerated distributed OPF & RNN/LSTM & Learning-assisted ADMM warm-start for distributed OPF convergence & Computational speed-up up to 250$\times$ & \cite{steven2026machine} \\
		\addlinespace[3.0pt]
		Fault and NTL detection & SVM, RF, SOM, K-means & AI-driven fault classification and anomaly detection using smart meter data & Detection accuracy exceeding 95\% & \cite{barja2021artificial} \\
		\bottomrule
	\end{tabular}
\end{table*}

Table \ref{tab:ai_power_systems} summarizes representative operation-level studies. The reported outcomes should be interpreted as task-specific results from the cited works rather than as directly comparable benchmarks, because the studies differ in network model, data availability, disturbance set, DER penetration, and validation environment.

Noise-trained generative adversarial networks (GANs) address these issues by learning mappings from local measurements (e.g., smart meters, PV and battery inverters) to system states (voltage magnitudes and phase angles). By incorporating Gaussian noise and simulating missing or corrupted data during training, GANs achieve strong robustness and outperform WLS and conventional bad-data detection methods, even under $\sim$30\% data interference \cite{liu2025noise}.

Similarly, deep learning models such as LSTM and attention-based recurrent networks generate pseudo-measurements, improving observability in systems with limited measurement availability \cite{shang2021achieving}. Their ability to capture nonlinear and temporal dynamics makes them well-suited for high-DER networks with reverse power flows and rapid voltage variations. Nevertheless, DSSE remains a high-risk deployment task because the estimator must remain credible under topology changes, bad data, communication delays, and unseen operating conditions.

\begin{figure*}[!t]
\centering
\includegraphics[width=0.70\textwidth]{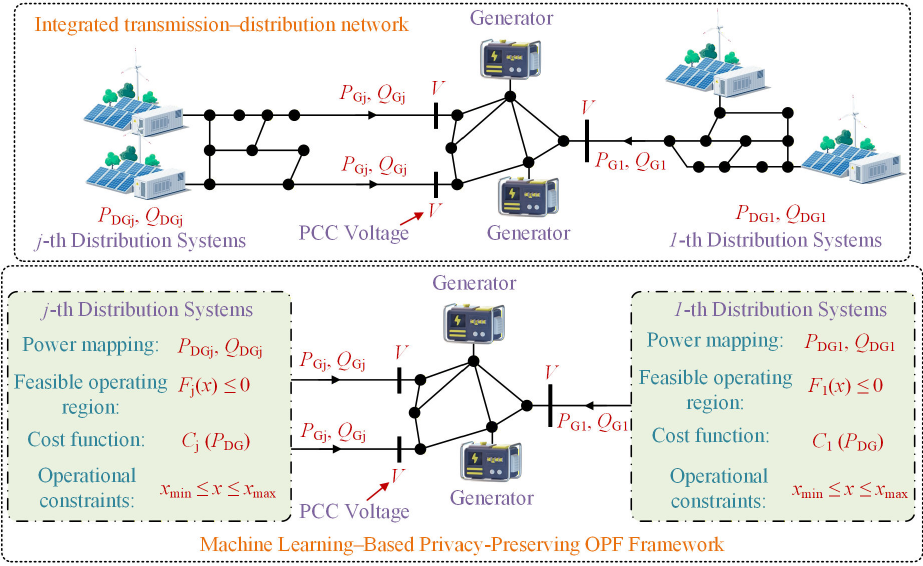}
\caption{Integrated transmission-distribution system with an ML-based privacy-preserving representation of distribution systems for OPF \cite{dindar2026privacy}.}
\label{fig:tso_dso}
\end{figure*}

\subsubsection{Topology Identification and Observability} 
\indent Distribution network topology is often uncertain or frequently changing due to switching operations and remains a critical input for reliable operation. AI techniques enable topology identification from measurement data, where voltage correlation analysis detects phase connectivity and transformer-customer relationships, and classifiers such as decision trees, multilayer perceptrons (MLPs), and logistic regression predict switch states \cite{xiong2022two}. Additionally, the alternating direction method of multipliers (ADMM) supports distributed topology estimation in large-scale systems.

In parallel, observability is enhanced by phasor measurement units (PMUs) and advanced metering infrastructure (AMI); however, high PMU costs limit full deployment \cite{baldwin2002power}. AI-based approaches, including deep neural networks (DNNs) and Bayesian methods, provide accurate state estimation with incomplete measurements \cite{jeyaraj2022optimum}. These methods improve situational awareness and operational reliability in partially observable networks. In particular, DNNs reduce the computational burden of Bayesian methods, but adapting to frequent topology changes remains a key challenge.

\subsubsection{Voltage and VAR Control with High DER Penetration} 
\indent High PV penetration in distribution networks causes voltage rise and fluctuations, requiring fast and coordinated reactive power control \cite{karimi2016photovoltaic}. DRL provides a model-free solution for volt-VAR control, but multi-agent setups face non-stationarity and uneven learning. To address this, graph meta-reinforcement learning (MLG-MATD3) integrates graph convolutional networks for topology awareness, meta-learning to predict agent actions, and self-attention in the critic for coordination. This approach outperforms MATD3, MADDPG, and MPC on IEEE 33-, 141-, and 322-bus systems, maintaining voltage violations below 1\% \cite{ge2025autonomous}.

Other DRL methods, including MADDPG and independent PPO, also enable decentralized voltage control but involve trade-offs in communication overhead, convergence stability, and scalability \cite{shi2023coordinated}. These methods may struggle with coordination as the number of agents increases or network conditions vary. To overcome these limitations, physics-informed and graph-based DRL approaches incorporate network topology and power flow constraints directly into the learning process \cite{cao2023physics}. This integration enhances robustness, reduces reliance on large training datasets, and improves generalization across different network sizes, operating conditions, and DER penetration levels.

\subsubsection{Privacy-Preserving TSO-DSO Coordination} 
\indent Effective TSO-DSO coordination is essential for utilizing distribution flexibility, but sharing sensitive information (e.g., network topology, line parameters, and customer load profiles) remains a key barrier. A machine learning-based privacy-preserving approach addresses this by training neural network models using only non-sensitive data, such as DG active/reactive power outputs and PCC voltage magnitudes \cite{dindar2026privacy}. As illustrated in Fig. \ref{fig:tso_dso}, the framework uses ML-based models to approximate the feasible operating region and map DG dispatch to PCC power flows. This enables the TSO to solve a single-round optimal power flow (OPF) problem and directly obtain DG setpoints without iterative data exchange.

The approach supports meshed distribution systems with multiple PCCs and non-rectangular PQ capability regions, overcoming key limitations of conventional aggregation methods. Case studies on a modified IEEE 30-bus transmission system integrated with multiple IEEE 33-bus distribution networks demonstrate near-optimal performance, with cost deviations around 0.36\%, full feasibility, and negligible computational overhead ($\sim$0.1s compared to AC-OPF) \cite{dindar2026privacy}. The method scales well with system size and operating conditions while maintaining accuracy. Importantly, it follows a ``privacy by design'' principle, as the ML models are trained on inherently non-sensitive datasets, making them resistant to reverse-engineering and ensuring secure coordination.

\begin{figure*}[!t]
\centering
\includegraphics[width=0.70\textwidth]{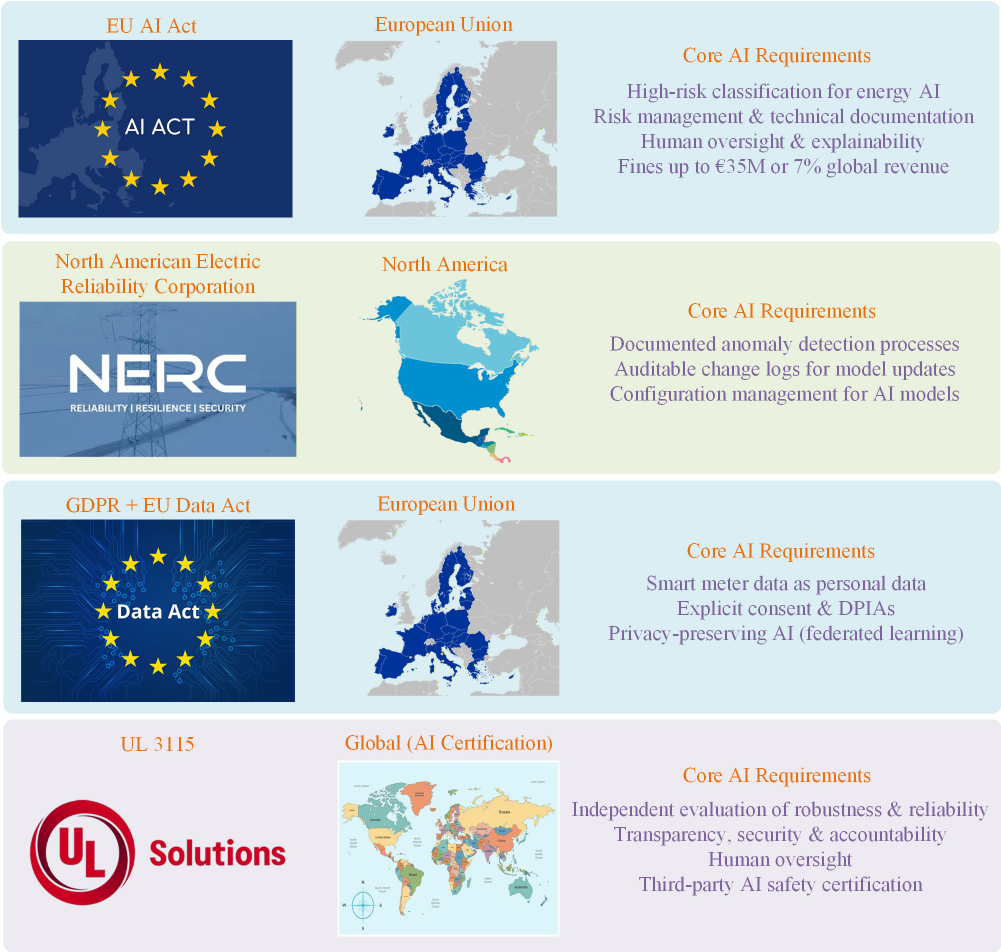}
\caption{Major AI compliance standards and regulatory requirements for power-converter-rich electrical systems.}
\label{fig:compliance}
\end{figure*}

\section{AI Compliance Standards}
As AI becomes integrated into grid control, energy management, cybersecurity monitoring, and customer-facing services, compliance becomes part of the engineering problem rather than a separate legal afterthought. Unlike rule-based automation, DL and RL models may change after retraining, rely on opaque internal representations, and interact with sensitive operational or customer data. These properties create requirements for traceability, documentation, human oversight, data governance, cybersecurity controls, and post-deployment monitoring. Fig. \ref{fig:compliance} summarizes the main compliance dimensions considered in this review.

\subsection{European Union AI Act and Energy Sector Compliance} 
\indent The European Union AI Act introduces a risk-based framework for AI systems, and energy-sector applications may fall into high-risk categories when they affect safety, critical infrastructure, or essential services \cite{cunha2025unlocking}. For power-converter-rich electrical systems, the engineering implication is that AI functions used for grid control, operational decision support, or infrastructure security should be designed with risk management, data quality, technical documentation, human oversight, transparency, and post-deployment monitoring in mind.

Implementation remains challenging because technical standards, conformity-assessment practices, and sector-specific interpretations continue to evolve. This uncertainty increases compliance complexity, particularly for smaller utilities and technology providers with limited regulatory resources. A practical response is compliance-by-design: model documentation, data provenance, version control, validation records, fallback modes, and human approval procedures should be prepared together with the AI function rather than added after deployment.

\subsection{Cybersecurity Compliance: NERC CIP Framework}
\indent In North America, the NERC Critical Infrastructure Protection (CIP) standards form the core cybersecurity framework for the Bulk Electric System (BES) \cite{rahman2025us}. For AI-enabled monitoring and control, the key engineering issue is auditability. DL-based intrusion detection, anomaly detection, or operational-security monitoring may identify abnormal patterns that rule-based tools miss, but their decisions must still be logged, reviewed, and explained well enough to support compliance evidence and incident response.

Cybersecurity compliance also affects model maintenance. AI-driven applications such as voltage control, state estimation, DER dispatch, and cyber-event detection require documented model updates, retraining triggers, configuration changes, access control, and supply-chain management \cite{sarkar2023study}. While AI can improve monitoring and threat detection, it can also introduce new attack surfaces through data poisoning, adversarial inputs, compromised edge devices, or unapproved model changes. Secure deployment therefore requires both cyber-physical robustness and compliance-grade change management.

\subsection{Data Privacy Compliance: GDPR, Data Act, and Privacy-Preserving AI} 
\indent AI deployment in power systems relies heavily on consumer and operational data, including high-resolution smart-meter readings that may reveal occupancy, appliance use, and behavioral patterns \cite{wang2024live}. Privacy requirements therefore constrain centralized data collection and model training. Methods such as federated learning, differential privacy, secure aggregation, and privacy-preserving feature design can reduce raw-data exposure while still supporting load forecasting, anomaly detection, and demand-side management \cite{badr2023privacy}.

Data-access rules for connected devices also affect AI services that use smart meters, grid sensors, virtual power plants, and energy-trading platforms \cite{jorgensen2025impact}. From an engineering perspective, these requirements imply that AI workflows should record data origin, user permissions, retention policies, aggregation level, model-update paths, and re-identification risks. Techniques such as differential privacy can help limit the exposure of consumption patterns while maintaining operational accuracy \cite{lian2025secure}. Together, privacy and data-access requirements shape how AI models can be trained, updated, and audited in customer-facing power-system applications.

\subsection{Emerging AI-Specific Certification: UL 3115} 
\indent UL 3115 represents an emerging attempt to evaluate the safety of AI systems in operational power and energy contexts \cite{UL3115_AI_Safety}. Its relevance to power-converter-rich electrical systems lies in the shift from algorithm accuracy toward system-level trustworthiness. Certification-oriented evaluation must consider data quality, model performance, failure handling, operational safeguards, robustness, reliability, transparency, security, accountability, and human oversight.

For the power sector, AI-specific certification can complement cybersecurity, privacy, and critical-infrastructure regulation by translating abstract trustworthiness requirements into auditable engineering evidence. Such evidence may include training-data records, model-version histories, stress-test results, failure-mode analysis, fallback control logic, human-override procedures, and monitoring logs. This direction is still developing, but it points toward a future in which AI-enabled forecasting, optimization, and autonomous control must be justified through a safety case rather than through performance metrics alone.

\section{Conclusion}
This paper reviewed AI for power-converter-rich electrical systems through a life-cycle lens spanning converter design, real-time control, system-level operation, and compliance-oriented governance. Across these stages, the surveyed literature shows a consistent shift from isolated data-driven tools toward hybrid workflows in which AI supports physics-based modeling, optimization, control, monitoring, and documentation. In design, surrogate modeling and AI-assisted synthesis can accelerate multi-physics exploration when they remain coupled to simulation and experimental verification. In control, supervised learning, DRL, and learning-augmented predictive control are most credible when their actions are bounded by constraints, safety filters, or stability-related evidence. In operation, edge AI, federated and graph learning, and privacy-preserving coordination are most relevant when they address converter-interfaced DERs, distribution observability, cyber-physical resilience, and condition monitoring.

Beyond reporting algorithmic performance, the review emphasized engineering constraints that determine deployability in safety-critical infrastructure. Key open challenges include stability and constraint certification for learning-based controllers, interpretability and out-of-distribution generalization of black-box models, data efficiency and sim-to-real transfer, microsecond-level inference latency on embedded hardware, robustness against electromagnetic interference and adversarial manipulation, and end-to-end governance of data and model updates.

The main implication is that trustworthy AI in this field should be evaluated by deployment readiness, not by accuracy, reward, or computation time alone. Future progress will depend on physics-integrated and uncertainty-aware learning, hardware-software co-design for real-time implementation, privacy-preserving distributed intelligence, and standardized evaluation protocols that connect benchmarks to grid codes, cybersecurity requirements, and certification-oriented safety cases. The resulting roadmap is strongest where direct converter-level validation exists, and it remains incomplete for large-scale field operation, long-term aging behavior, and evolving compliance regimes. Recognizing this boundary is essential for translating AI advances into reliable, secure, and auditable capabilities for next-generation converter-rich power systems.




\begin{thebibliography}{100}
\providecommand{\url}[1]{#1}
\csname url@samestyle\endcsname
\providecommand{\newblock}{\relax}
\providecommand{\bibinfo}[2]{#2}
\providecommand{\BIBentrySTDinterwordspacing}{\spaceskip=0pt\relax}
\providecommand{\BIBentryALTinterwordstretchfactor}{4}
\providecommand{\BIBentryALTinterwordspacing}{\spaceskip=\fontdimen2\font plus
\BIBentryALTinterwordstretchfactor\fontdimen3\font minus
  \fontdimen4\font\relax}
\providecommand{\BIBforeignlanguage}[2]{{%
\expandafter\ifx\csname l@#1\endcsname\relax
\typeout{** WARNING: IEEEtran.bst: No hyphenation pattern has been}%
\typeout{** loaded for the language `#1'. Using the pattern for}%
\typeout{** the default language instead.}%
\else
\language=\csname l@#1\endcsname
\fi
#2}}
\providecommand{\BIBdecl}{\relax}
\BIBdecl

\bibitem{TPEL.2026.3675661}
F.~Blaabjerg, K.~Zhang, Y.~Song, and Y.~Zhang, ``Advances in reliability and
  artificial intelligence for power electronic systems,'' \emph{IEEE
  Transactions on Power Electronics}, vol.~41,
  DOI
  10.1109/TPEL.2026.3675661, no.~3, pp. 2890--2905, 2026.

\bibitem{TIA.2025.3529797}
S.~Subedi, Y.~Gui, and Y.~Xue, ``Applications of data-driven dynamic modeling
  of power converters in power systems: An overview,'' \emph{IEEE Transactions
  on Industry Applications}, vol.~61,
  DOI
  10.1109/TIA.2025.3529797, no.~1, pp. 112--125, 2025.

\bibitem{TPEL.2020.3024914}
S.~Zhao, F.~Blaabjerg, and H.~Wang, ``An overview of artificial intelligence
  applications for power electronics,'' \emph{IEEE Transactions on Power
  Electronics}, vol.~68, DOI
  10.1109/TPEL.2020.3024914, no.~4, pp. 3029--3045, 2021.

\bibitem{RSER.2025.116591}
J.~Ye \emph{et~al.}, ``An overview of reinforcement learning for power
  electronic converters: Topology derivation, parameter design, and control
  implementation,'' \emph{Renewable and Sustainable Energy Reviews}, vol. 228,
  DOI
  10.1016/j.rser.2025.116591, p. 116591, 2026.

\bibitem{APENERGY.2025.126923}
A.~Safari, A.~Oshnoei, and F.~Blaabjerg, ``A review of recent ai applications
  in next-generation power electronics,'' \emph{Applied Energy}, vol. 402,
  DOI
  10.1016/j.apenergy.2025.126923, p. 126923, 2025.

\bibitem{CSEEJPES.2020.02700}
M.~Khodayar, G.~Liu, J.~Wang, and M.~E. Khodayar, ``Deep learning in power
  systems research: A review,'' \emph{CSEE Journal of Power and Energy
  Systems}, vol.~7, DOI
  10.17775/CSEEJPES.2020.02700, no.~2, pp. 209--220, 2021.

\bibitem{JESTIE.2022.3198504}
E.~Mohammadi \emph{et~al.}, ``A review on application of artificial
  intelligence techniques in microgrids,'' \emph{IEEE Journal of Emerging and
  Selected Topics in Industrial Electronics}, vol.~3,
  DOI
  10.1109/JESTIE.2022.3198504, no.~4, pp. 878--890, 2022.

\bibitem{ICCAD.2021.9643548}
S.~Fan \emph{et~al.}, ``From specification to topology: Automatic power
  converter design via reinforcement learning,'' in \emph{Proc. IEEE/ACM
  International Conference on Computer-Aided Design (ICCAD)},
  DOI
  10.1109/ICCAD51958.2021.9643548, pp. 1--9, 2021.

\bibitem{TPWRS.2022.3162473}
B.~Huang and J.~Wang, ``Applications of physics-informed neural networks in
  power systems - a review,'' \emph{IEEE Transactions on Power Systems},
  vol.~38, DOI
  10.1109/TPWRS.2022.3162473, no.~1, pp. 572--588, 2023.

\bibitem{TIA.2025.3626472}
A.~Selim \emph{et~al.}, ``Safe deep reinforcement learning for robust frequency
  and voltage-constrained networked microgrid restoration,'' \emph{IEEE
  Transactions on Industry Applications}, vol.~62,
  DOI
  10.1109/TIA.2025.3626472, no.~2, pp. 3635--3647, 2026.

\bibitem{OJPEL.2025.3619673}
A.~Rajamallaiah \emph{et~al.}, ``Deep reinforcement learning for power
  converter control: A comprehensive review of applications and challenges,''
  \emph{IEEE Open Journal of Power Electronics}, vol.~13,
  DOI
  10.1109/OJPEL.2025.3619673, pp. 111\,976--111\,995, 2025.

\bibitem{TPEL.2024.3358912}
Z.~Feng \emph{et~al.}, ``Deep reinforcement learning assisted hybrid
  five-variable modulation scheme for dab converters to reduce rms current and
  expand zvs operation,'' \emph{IEEE Transactions on Power Electronics},
  vol.~39, DOI
  10.1109/TPEL.2024.3358912, no.~7, pp. 8114--8128, 2024.

\bibitem{OJIA.2023.3338534}
Y.~Gao, S.~Wang, T.~Dragicevic, P.~Wheeler, and P.~Zanchetta, ``Artificial
  intelligence techniques for enhancing the performance of controllers in power
  converter-based systems---an overview,'' \emph{IEEE Open Journal of Industry
  Applications}, vol.~4, DOI
  10.1109/OJIA.2023.3338534, pp. 385--403, 2023.

\bibitem{MIE.2022.3211125}
D.~Gebbran \emph{et~al.}, ``Cloud and edge computing for smart management of
  power electronic converter fleets: A key connective fabric to enable the
  green transition,'' \emph{IEEE Industrial Electronics Magazine}, vol.~17,
  DOI
  10.1109/MIE.2022.3211125, no.~2, pp. 6--19, 2022.

\bibitem{TPWRS.2020.3001919}
X.~Lei, Z.~Yang, J.~Yu, J.~Zhao, Q.~Gao, and H.~Yu, ``Data-driven optimal power
  flow: A physics-informed machine learning approach,'' \emph{IEEE Transactions
  on Power Systems}, vol.~36,
  DOI
  10.1109/TPWRS.2020.3001919, no.~1, pp. 346--354, 2021.

\bibitem{TII.2022.3163137}
L.~Lv, Z.~Wu, L.~Zhang, B.~B. Gupta, and Z.~Tian, ``An edge-ai based
  forecasting approach for improving smart microgrid efficiency,'' \emph{IEEE
  Transactions on Industrial Informatics}, vol.~18,
  DOI
  10.1109/TII.2022.3163137, no.~11, pp. 7946--7954, 2022.

\bibitem{TSG.2024.3466768}
M.~Sayak \emph{et~al.}, ``Resilient control of networked microgrids using
  vertical federated reinforcement learning: Designs and real-time test-bed
  validations,'' \emph{IEEE Transactions on Smart Grid}, vol.~16,
  DOI
  10.1109/TSG.2024.3466768, no.~2, pp. 1897--1910, 2025.

\bibitem{TNNLS.2022.3232630}
Y.~Li, S.~He, Y.~Li, Y.~Shi, and Z.~Zeng, ``Federated multiagent deep
  reinforcement learning approach via physics-informed reward for
  multimicrogrid energy management,'' \emph{IEEE Transactions on Neural
  Networks and Learning Systems}, vol.~35,
  DOI
  10.1109/TNNLS.2022.3232630, no.~5, pp. 5902--5915, 2024.

\bibitem{APENERGY.2025.126900}
P.~Lu \emph{et~al.}, ``Hierarchical reserve-based distributionally robust
  chance-constrained optimization for integrated electricity-heat systems
  during cold waves,'' \emph{Applied Energy}, vol. 402,
  DOI
  10.1016/j.apenergy.2025.126900, p. 126900, 2026.

\bibitem{ACCESS.2021.3131502}
A.~Kumar, M.~Alaraj, M.~Rizwan, and U.~Nangia, ``Novel ai based energy
  management system for smart grid with res integration,'' \emph{IEEE Access},
  vol.~9, DOI
  10.1109/ACCESS.2021.3131502, pp. 162\,530--162\,542, 2021.

\bibitem{TDSC.2021.3118636}
M.~Farajzadeh-Zanjani, E.~Hallaji, R.~Razavi-Far, and M.~Saif,
  ``Generative-adversarial class-imbalance learning for classifying
  cyber-attacks and faults - a cyber-physical power system,'' \emph{IEEE
  Transactions on Dependable and Secure Computing}, vol.~19,
  DOI
  10.1109/TDSC.2021.3118636, no.~6, pp. 4068--4081, 2022.

\bibitem{APEC.2022.9773482}
S.~Zhao \emph{et~al.}, ``Physics-informed machine learning for parameter
  estimation of dc-dc converter,'' in \emph{2022 IEEE Applied Power Electronics
  Conference and Exposition (APEC)},
  DOI
  10.1109/APEC43599.2022.9773482, 2022.

\bibitem{ACCESS.2023.3248511}
S.~Ahmad, M.~Shafiullah, C.~B. Ahmed, and M.~Alowaifeer, ``A review of
  microgrid energy management and control strategies,'' \emph{IEEE Access},
  vol.~11, DOI
  10.1109/ACCESS.2023.3248511, pp. 21\,733--21\,757, 2023.

\bibitem{JAS.2023.123657}
A.~Joshi, S.~Capezza, A.~Alhaji, and M.-Y. Chow, ``Survey on ai and machine
  learning techniques for microgrid energy management systems,'' \emph{IEEE/CAA
  Journal of Automatica Sinica}, vol.~10,
  DOI
  10.1109/JAS.2023.123657, no.~7, pp. 1513--1529, 2023.

\bibitem{fang2018review}
X.~Fang, S.~Lin, X.~Huang, F.~Lin, Z.~Yang, and S.~Igarashi, ``A review of
  data-driven prognostic for {IGBT} remaining useful life,'' \emph{IEEE
  Transactions on Power Electronics}, vol.~33,
  DOI
  10.1109/TPEL.2017.2785441, no.~11, pp. 9844--9855, 2018.

\bibitem{yu2022congo}
J.~Yu, A.~Alqahtani, and A.~P. Meliopoulos, ``{CONGO}: Scalable online anomaly
  detection and localization in power electronics networks,'' \emph{IEEE
  Internet of Things Journal}, vol.~9,
  DOI
  10.1109/JIOT.2022.3143123, no.~16, pp. 14\,754--14\,765, 2022.

\bibitem{cui2017svm}
J.~Cui, G.~Shi, and C.~Gong, ``A fast classification method of faults in power
  electronic circuits based on support vector machines,'' \emph{Metrology and
  Measurement Systems}, vol.~24,
  DOI 10.1515/mms-2017-0056,
  no.~4, pp. 701--720, 2017.

\bibitem{ren2024transfer}
C.~Ren, T.~Wang, Z.~Y. Dong, and R.~Zhang, ``Transfer extreme learning machine
  for power system cross-fault and cross-scale stability assessment with
  limited guide instances,'' \emph{IEEE Transactions on Power Systems},
  vol.~39, DOI
  10.1109/TPWRS.2024.3366433, no.~3, pp. 5431--5434, 2024.

\bibitem{diao2009decision}
R.~Diao, K.~Sun, V.~Vittal, R.~J. O'Keefe, M.~R. Richardson, N.~Bhatt,
  D.~Stradford, and S.~K. Sarawgi, ``Decision tree-based online voltage
  security assessment using {PMU} measurements,'' \emph{IEEE Transactions on
  Power Systems}, vol.~24,
  DOI
  10.1109/TPWRS.2009.2016528, no.~2, pp. 832--839, 2009.

\bibitem{patel2008maximum}
H.~Patel and V.~Agarwal, ``Maximum power point tracking scheme for {PV} systems
  operating under partially shaded conditions,'' \emph{IEEE Transactions on
  Industrial Electronics}, vol.~55,
  DOI
  10.1109/TIE.2008.917118, no.~4, pp. 1689--1698, 2008.

\bibitem{hochreiter1997lstm}
S.~Hochreiter and J.~Schmidhuber, ``Long short-term memory,'' \emph{Neural
  Computation}, vol.~9, DOI
  10.1162/neco.1997.9.8.1735, no.~8, pp. 1735--1780, 1997.

\bibitem{cho2014gru}
K.~Cho, B.~van Merri{\"e}nboer, {\c{C}}.~G{\"u}l{\c{c}}ehre, D.~Bahdanau,
  F.~Bougares, H.~Schwenk, and Y.~Bengio, ``Learning phrase representations
  using {RNN} encoder--decoder for statistical machine translation,'' in
  \emph{Proceedings of the 2014 Conference on Empirical Methods in Natural
  Language Processing (EMNLP)},
  DOI 10.3115/v1/D14-1179, pp.
  1724--1734.\hskip 1em plus 0.5em minus 0.4em\relax Doha, Qatar: Association
  for Computational Linguistics, 2014.

\bibitem{xue2020lstmfault}
Z.~Y. Xue, K.~S. Xiahou, M.~S. Li, T.~Y. Ji, and Q.~H. Wu, ``Diagnosis of
  multiple open-circuit switch faults based on long short-term memory network
  for {DFIG}-based wind turbine systems,'' \emph{IEEE Journal of Emerging and
  Selected Topics in Power Electronics}, vol.~8,
  DOI
  10.1109/JESTPE.2019.2908981, no.~3, pp. 2600--2610, 2020.

\bibitem{krizhevsky2012cnn}
A.~Krizhevsky, I.~Sutskever, and G.~E. Hinton, ``Imagenet classification with
  deep convolutional neural networks,'' in \emph{Advances in Neural Information
  Processing Systems 25}, pp. 1097--1105, 2012.

\bibitem{hang2023cnnfault}
J.~Hang, X.~Shu, S.~Ding, and Y.~Huang, ``Robust open-circuit fault diagnosis
  for {PMSM} drives using wavelet convolutional neural network with small
  samples of normalized current vector trajectory graph,'' \emph{IEEE
  Transactions on Industrial Electronics}, vol.~70,
  DOI
  10.1109/TIE.2022.3231304, no.~8, pp. 7653--7663, 2023.

\bibitem{vaswani2017transformer}
A.~Vaswani, N.~Shazeer, N.~Parmar, J.~Uszkoreit, L.~Jones, A.~N. Gomez,
  {\L}.~Kaiser, and I.~Polosukhin, ``Attention is all you need,'' in
  \emph{Advances in Neural Information Processing Systems 30}, pp. 5998--6008,
  2017.

\bibitem{liao2022gnn}
W.~Liao, B.~Bak-Jensen, J.~R. Pillai, Y.~Wang, and Y.~Wang, ``A review of graph
  neural networks and their applications in power systems,'' \emph{Journal of
  Modern Power Systems and Clean Energy}, vol.~10,
  DOI
  10.35833/MPCE.2021.000058, no.~2, pp. 345--360, 2022.

\bibitem{sutton2018rlintro}
R.~S. Sutton and A.~G. Barto, \emph{Reinforcement Learning: An Introduction},
  2nd~ed.\hskip 1em plus 0.5em minus 0.4em\relax Cambridge, MA: MIT Press,
  2018.

\bibitem{ye2026rloverview}
J.~Ye, W.~Xuan, Q.~Guo, Y.~Liu, B.~Wang, X.~Zhang, and H.~H.~C. Iu, ``An
  overview of reinforcement learning for power electronic converters: Topology
  derivation, parameter design, and control implementation,'' \emph{Renewable
  and Sustainable Energy Reviews}, vol. 228,
  DOI
  10.1016/j.rser.2025.116591, p. 116591, 2026.

\bibitem{gao2023aicontrol}
Y.~Gao, S.~Wang, T.~Dragicevic, P.~Wheeler, and P.~Zanchetta, ``Artificial
  intelligence techniques for enhancing the performance of controllers in power
  converter-based system: An overview,'' \emph{IEEE Open Journal of Industry
  Applications}, vol.~4, DOI
  10.1109/OJIA.2023.3338534, pp. 366--375, 2023.

\bibitem{mnih2015dqn}
V.~Mnih, K.~Kavukcuoglu, D.~Silver, A.~A. Rusu, J.~Veness, M.~G. Bellemare,
  A.~Graves, M.~Riedmiller, A.~K. Fidjeland, G.~Ostrovski, S.~Petersen,
  C.~Beattie, A.~Sadik, I.~Antonoglou, H.~King, D.~Kumaran, D.~Wierstra,
  S.~Legg, and D.~Hassabis, ``Human-level control through deep reinforcement
  learning,'' \emph{Nature}, vol. 518,
  DOI 10.1038/nature14236, no.
  7540, pp. 529--533, 2015.

\bibitem{lillicrap2015ddpg}
T.~P. Lillicrap, J.~J. Hunt, A.~Pritzel, N.~Heess, T.~Erez, Y.~Tassa,
  D.~Silver, and D.~Wierstra, ``Continuous control with deep reinforcement
  learning,'' \emph{arXiv preprint arXiv:1509.02971},
  DOI
  10.48550/arXiv.1509.02971, 2015.

\bibitem{haarnoja2018sac}
T.~Haarnoja, A.~Zhou, P.~Abbeel, and S.~Levine, ``Soft actor-critic: Off-policy
  maximum entropy deep reinforcement learning with a stochastic actor,'' in
  \emph{Proceedings of the 35th International Conference on Machine Learning},
  ser. Proceedings of Machine Learning Research, vol.~80, pp. 1861--1870.\hskip
  1em plus 0.5em minus 0.4em\relax PMLR, 2018.

\bibitem{brown2020gpt3}
T.~B. Brown, B.~Mann, N.~Ryder, M.~Subbiah, J.~Kaplan, P.~Dhariwal,
  A.~Neelakantan, P.~Shyam, G.~Sastry, A.~Askell, S.~Agarwal, A.~Herbert-Voss,
  G.~Krueger, T.~Henighan, R.~Child, A.~Ramesh, D.~M. Ziegler, J.~Wu,
  C.~Winter, C.~Hesse, M.~Chen, E.~Sigler, M.~Litwin, S.~Gray, B.~Chess,
  J.~Clark, C.~Berner, S.~McCandlish, A.~Radford, I.~Sutskever, and D.~Amodei,
  ``Language models are few-shot learners,'' in \emph{Advances in Neural
  Information Processing Systems 33},
  DOI
  10.5555/3495724.3495883, pp. 1877--1901, 2020.

\bibitem{raffel2020t5}
\BIBentryALTinterwordspacing
C.~Raffel, N.~Shazeer, A.~Roberts, K.~Lee, S.~Narang, M.~Matena, Y.~Zhou,
  W.~Li, and P.~J. Liu, ``Exploring the limits of transfer learning with a
  unified text-to-text transformer,'' \emph{Journal of Machine Learning
  Research}, vol.~21, no. 140, pp. 1--67, 2020. [Online]. Available:
  \arxivurl{https://jmlr.org/papers/v21/20-074.html}
\BIBentrySTDinterwordspacing

\bibitem{ouyang2022instructgpt}
L.~Ouyang, J.~Wu, X.~Jiang, D.~Almeida, C.~Wainwright, P.~Mishkin, C.~Zhang,
  S.~Agarwal, K.~Slama, A.~Ray, J.~Schulman, J.~Hilton, F.~Kelton, L.~Miller,
  M.~Simens, A.~Askell, P.~Welinder, P.~Christiano, J.~Leike, and R.~Lowe,
  ``Training language models to follow instructions with human feedback,'' in
  \emph{Advances in Neural Information Processing Systems 35},
  DOI
  10.5555/3600270.3602281, pp. 27\,730--27\,744, 2022.

\bibitem{lewis2020rag}
P.~Lewis, E.~Perez, A.~Piktus, F.~Petroni, V.~Karpukhin, N.~Goyal,
  H.~K{\"u}ttler, M.~Lewis, W.-t. Yih, T.~Rockt{\"a}schel, S.~Riedel, and
  D.~Kiela, ``Retrieval-augmented generation for knowledge-intensive nlp
  tasks,'' in \emph{Advances in Neural Information Processing Systems 33},
  DOI
  10.5555/3495724.3496517, pp. 9459--9474, 2020.

\bibitem{abouali2026agenticsurvey}
M.~Abou~Ali, F.~Dornaika, and J.~Charafeddine, ``Agentic ai: a comprehensive
  survey of architectures, applications, and future directions,''
  \emph{Artificial Intelligence Review}, vol.~59,
  DOI
  10.1007/s10462-025-11422-4, no.~11, 2026.

\bibitem{yao2023react}
\BIBentryALTinterwordspacing
S.~Yao, J.~Zhao, D.~Yu, N.~Du, I.~Shafran, K.~R. Narasimhan, and Y.~Cao,
  ``React: Synergizing reasoning and acting in language models,'' in \emph{The
  Eleventh International Conference on Learning Representations (ICLR)},
  DOI
  10.48550/arXiv.2210.03629, 2023. [Online]. Available:
  \arxivurl{https://openreview.net/forum?id=WE_vluYUL-X}
\BIBentrySTDinterwordspacing

\bibitem{schick2023toolformer}
T.~Schick, J.~Dwivedi-Yu, R.~Dessi, R.~Raileanu, M.~Lomeli, E.~Hambro,
  L.~Zettlemoyer, N.~Cancedda, and T.~Scialom, ``Toolformer: Language models
  can teach themselves to use tools,'' in \emph{Advances in Neural Information
  Processing Systems 36}, DOI
  10.5555/3666122.3668384, 2023.

\bibitem{park2023generativeagents}
J.~S. Park, J.~C. O'Brien, C.~J. Cai, M.~R. Morris, P.~Liang, and M.~S.
  Bernstein, ``Generative agents: Interactive simulacra of human behavior,'' in
  \emph{UIST '23: Proceedings of the 36th Annual ACM Symposium on User
  Interface Software and Technology},
  DOI
  10.1145/3586183.3606763, pp. 1--22, 2023.

\bibitem{wu2024autogen}
\BIBentryALTinterwordspacing
Q.~Wu, G.~Bansal, J.~Zhang, Y.~Wu, B.~Li, E.~Zhu, L.~Jiang, X.~Zhang, S.~Zhang,
  J.~Liu, A.~H. Awadallah, R.~W. White, D.~Burger, and C.~Wang, ``Autogen:
  Enabling next-gen llm applications via multi-agent conversations,'' in
  \emph{Proceedings of the First Conference on Language Modeling (COLM)},
  DOI
  10.48550/arXiv.2308.08155, 2024. [Online]. Available:
  \arxivurl{https://openreview.net/forum?id=BAakY1hNKS}
\BIBentrySTDinterwordspacing

\bibitem{FraunhoferPINN}
{Fraunhofer Institute}, ``Physics-informed neural networks for magnetostatic
  problems on axisymmetric transformer geometries,'' 2024, fraunhofer Publica.

\bibitem{Chen2025Probabilistic}
X.~Chen \emph{et~al.}, ``Optimal parameter design for power electronic
  converters using a probabilistic learning-based stochastic surrogate model,''
  \emph{Next Energy}, vol.~9,
  DOI
  10.1016/j.nenergy.2025.100464, p. 100464, 2025.

\bibitem{PowerModuleAI}
Anonymous, ``Optimizing chip area in power module design: Comparison of
  traditional and {AI} surrogate models for thermal resistance calculation,''
  in \emph{VDE Conference Proceedings}.\hskip 1em plus 0.5em minus 0.4em\relax
  VDE Verlag, 2025.

\bibitem{Viarouge2024}
I.~Viarouge \emph{et~al.}, ``Machine learning-based design tool for magnetic
  components in power electronics,'' in \emph{Proceedings of IPAC'23}, 2024,
  jACoW Publishing.

\bibitem{Lin2025}
F.~Lin, X.~Li, W.~Lei, J.~J. Rodriguez-Andina, J.~M. Guerrero, C.~Wen,
  X.~Zhang, and H.~Ma, ``{PE-GPT}: A new paradigm for power electronics
  design,'' \emph{IEEE Transactions on Industrial Electronics}, vol.~72,
  DOI
  10.1109/TIE.2024.3454408, no.~4, pp. 3778--3791, 2025.

\bibitem{Wang2022}
S.~Wang, G.~Calderon-Lopez, and W.~Ming, ``Artificial neural networks-based
  multi-objective design methodology for wide-bandgap power electronics
  converters,'' \emph{IEEE Open Journal of Power Electronics}, vol.~3,
  DOI
  10.1109/OJPEL.2022.3204630, pp. 664--678, 2022.

\bibitem{Shen2024}
X.~Shen, S.~Zhao, H.~Wang, and F.~Blaabjerg, ``Artificial intelligence
  applications in high-frequency magnetic components design for power
  electronics systems: An overview,'' \emph{IEEE Transactions on Power
  Electronics}, vol.~39, DOI
  10.1109/TPEL.2024.3381431, no.~7, pp. 8478--8496, 2024.

\bibitem{Li2026}
Y.~Li \emph{et~al.}, ``Frequency-band-aware physics-informed generative
  adversarial network for {EMI} prediction and adaptive suppression in {SiC}
  power converters,'' \emph{Electronics}, vol.~15,
  DOI
  10.3390/electronics15081560, no.~8, p. 1560, 2026.

\bibitem{Tlig2026}
M.~Tlig, M.~Kadi, and Z.~Riah, ``Comparative study of {AI} methods for {EMC}
  prediction in power electronics applications,'' \emph{Electronics}, vol.~15,
  DOI
  10.3390/electronics15010165, no.~1, p. 165, 2026.

\bibitem{Lu2026}
M.~Lu, K.~Jia, R.~Goswami, and Y.~Hu, ``Intelligent self-tuning active {EMI}
  filtering for electrified automotive power systems using reinforcement
  learning,'' 2026, arXiv preprint arXiv:2604.28084.

\bibitem{Halidu2025}
A.~M. Halidu, S.~Nunoo, and J.~C. Attachie, ``{ANN} optimised {RPWM} technique
  for minimisation of conducted {EMI} in three-phase voltage source
  inverters,'' \emph{Power Electronics and Drives}, vol.~10,
  DOI 10.2478/pead-2025-0028,
  no.~1, pp. 406--423, 2025.

\bibitem{EMI_Mitigation_Toolset}
Anonymous, ``Electromagnetic interference mitigation toolset for power
  electronic systems,'' Master's thesis, University of Wisconsin--Madison,
  2025, available via Minds@UW.

\bibitem{Yang2025}
X.~Yang, Y.~Xiao, L.~Shu, W.~Taborsky, and D.~Yang, ``Automatic power
  electronic {PCB} layout design based on generative {AI}: The pathway towards
  next-generation hardware compiler,'' in \emph{2025 Energy Conversion Congress
  \& Expo Europe (ECCE Europe)}, pp. 1--5, 2025.

\bibitem{EMA_EDA_PCB_AI}
``{PCB} power design with {AI},'' EMA Design Automation, 2025, webinar;
  accessed 2025.

\bibitem{D2S_FLOW}
C.~Chen \emph{et~al.}, ``{D2S-FLOW}: Automated parameter extraction from
  datasheets for {SPICE} model generation using large language models,'' 2025,
  arXiv preprint arXiv:2502.16540.

\bibitem{Flux_AI_Tool}
``Engineers can ditch datasheets with {Flux} generative {AI} tool,'' EE Power,
  2023, online; accessed 2025.

\bibitem{Graphwise_Statnett}
``{Graphwise-Statnett} project: Democratizing the use of power system models,''
  Graphwise.ai, 2025, online; accessed 2025.

\bibitem{Nau2025}
S.~Nau, J.~Krummenauer, and A.~Zimmermann, ``Evaluating {LLM}-based workflows
  for switched-mode power supply design,'' 2025, arXiv preprint
  arXiv:2507.10639v2.

\bibitem{Chen2025Collaborative}
Y.~Chen, Y.~Shang, Y.~Mo, and X.~Jin, ``Intelligent power electronics design: A
  collaborative framework of deep reinforcement learning and large language
  models with applications,'' \emph{Journal of Electrical Engineering},
  vol.~20, DOI
  10.11985/2025.05.003, no.~5, pp. 24--34, 2025.

\bibitem{Poddar2026}
S.~Poddar \emph{et~al.}, ``{HeaRT}: A hierarchical circuit reasoning tree-based
  agentic framework for {AMS} design optimization,'' 2026, arXiv preprint
  arXiv:2511.19669v2.

\bibitem{Qualcomm_Allegro_AI}
``{PCB} design time savings with {AI}: {Qualcomm} case story,'' EMA Design
  Automation, 2025, online; accessed 2025.

\bibitem{Dragicevic2019}
T.~Dragicevic, P.~Wheeler, and F.~Blaabjerg, ``Artificial intelligence aided
  automated design for reliability of power electronic systems,'' \emph{IEEE
  Transactions on Power Electronics}, vol.~34,
  DOI
  10.1109/TPEL.2018.2883947, no.~8, pp. 7161--7171, 2019.

\bibitem{TPEL.2020.2977765}
M.~Hajihosseini, M.~Andalibi, M.~Gheisarnejad, H.~Farsizadeh, and M.-H.
  Khooban, ``Dc/dc power converter control-based deep machine learning
  techniques: Real-time implementation,'' \emph{IEEE Transactions on Power
  Electronics}, vol.~35, DOI
  10.1109/TPEL.2020.2977765, no.~10, pp. 9971--9977, 2020.

\bibitem{TII.2020.2969729}
S.~Lucia, D.~Navarro, B.~Karg, H.~Sarnago, and {\'O}.~Luc{\'i}a, ``Deep
  learning-based model predictive control for resonant power converters,''
  \emph{IEEE Transactions on Industrial Informatics}, vol.~17,
  DOI
  10.1109/TII.2020.2969729, no.~1, pp. 409--420, 2021.

\bibitem{APEC43599.2022.9773436}
K.~Yu, F.~Zhuo, F.~Wang, and X.~Jiang, ``Deep-learning-based steady-state
  modeling and model predictive control for cllc dc-dc resonant converter in dc
  distribution system,'' in \emph{2022 IEEE Applied Power Electronics
  Conference and Exposition (APEC)},
  DOI
  10.1109/APEC43599.2022.9773436, pp. 1--5, 2022.

\bibitem{TPEL.2022.3172681}
M.~Abu-Ali, F.~Berkel, M.~Manderla, S.~Reimann, R.~Kennel, and M.~Abdelrahem,
  ``Deep learning-based long-horizon mpc: Robust, high performing, and
  computationally efficient control for pmsm drives,'' \emph{IEEE Transactions
  on Power Electronics}, vol.~37,
  DOI
  10.1109/TPEL.2022.3172681, no.~10, pp. 12\,486--12\,501, 2022.

\bibitem{TIE.2020.3038064}
S.~Wang, T.~Dragicevic, G.~F. Gontijo, S.~K. Chaudhary, and R.~Teodorescu,
  ``Machine learning emulation of model predictive control for modular
  multilevel converters,'' \emph{IEEE Transactions on Industrial Electronics},
  vol.~68, DOI
  10.1109/TIE.2020.3038064, no.~11, pp. 11\,628--11\,634, 2021.

\bibitem{TIE.2021.3076721}
D.~Wang, Z.~J. Shen, X.~Yin, S.~Tang, X.~Liu, C.~Zhang, J.~Wang, J.~Rodriguez,
  and M.~Norambuena, ``Model predictive control using artificial neural network
  for power converters,'' \emph{IEEE Transactions on Industrial Electronics},
  vol.~69, DOI
  10.1109/TIE.2021.3076721, no.~4, pp. 3689--3699, 2022.

\bibitem{TIE.2022.3208594}
X.~Liu, L.~Qiu, Y.~Fang, and J.~Rodr{\'i}guez, ``Predictor-based data-driven
  model-free adaptive predictive control of power converters using machine
  learning,'' \emph{IEEE Transactions on Industrial Electronics}, vol.~70,
  DOI
  10.1109/TIE.2022.3208594, no.~8, pp. 7591--7603, 2023.

\bibitem{TIE.2023.3303646}
X.~Liu, L.~Qiu, Y.~Fang, K.~Wang, Y.~Li, and J.~Rodr{\'i}guez, ``Finite
  control-set learning predictive control for power converters,'' \emph{IEEE
  Transactions on Industrial Electronics}, vol.~71,
  DOI
  10.1109/TIE.2023.3303646, no.~7, pp. 8190--8196, 2024.

\bibitem{TPEL.2022.3194518}
W.~Wu, L.~Qiu, X.~Liu, F.~Guo, J.~Rodriguez, J.~Ma, and Y.~Fang, ``Data-driven
  iterative learning predictive control for power converters,'' \emph{IEEE
  Transactions on Power Electronics}, vol.~37,
  DOI
  10.1109/TPEL.2022.3194518, no.~12, pp. 14\,028--14\,033, 2022.

\bibitem{JESTPE.2026.3673944}
P.~Hui, C.~Cui, P.~Lin, A.~M. Y.~M. Ghias, X.~Niu, and C.~Zhang, ``A
  physics-informed imitation learning framework for adaptive control of power
  converters,'' \emph{IEEE Journal of Emerging and Selected Topics in Power
  Electronics}, DOI
  10.1109/JESTPE.2026.3673944, pp. 1--1, 2026.

\bibitem{TIE.2020.2969116}
M.~Novak and T.~Dragicevic, ``Supervised imitation learning of finite-set model
  predictive control systems for power electronics,'' \emph{IEEE Transactions
  on Industrial Electronics}, vol.~68,
  DOI
  10.1109/TIE.2020.2969116, no.~2, pp. 1717--1723, 2021.

\bibitem{TPWRS.2019.2941134}
J.~Duan, D.~Shi, R.~Diao, H.~Li, Z.~Wang, B.~Zhang, D.~Bian, and Z.~Yi,
  ``Deep-reinforcement-learning-based autonomous voltage control for power grid
  operations,'' \emph{IEEE Transactions on Power Systems}, vol.~35,
  DOI
  10.1109/TPWRS.2019.2941134, no.~1, pp. 814--817, 2020.

\bibitem{JESTPE.2022.3189078}
B.~Huangfu, C.~Cui, C.~Zhang, and L.~Xu, ``Learning-based optimal large-signal
  stabilization for dc/dc boost converters feeding cpls via deep reinforcement
  learning,'' \emph{IEEE Journal of Emerging and Selected Topics in Power
  Electronics}, vol.~11,
  DOI
  10.1109/JESTPE.2022.3189078, no.~6, pp. 5592--5601, 2023.

\bibitem{TIE.2022.3192676}
C.~Cui, T.~Yang, Y.~Dai, C.~Zhang, and Q.~Xu, ``Implementation of transferring
  reinforcement learning for dc--dc buck converter control via duty ratio
  mapping,'' \emph{IEEE Transactions on Industrial Electronics}, vol.~70,
  DOI
  10.1109/TIE.2022.3192676, no.~6, pp. 6141--6150, 2023.

\bibitem{TIE.2025.3572981}
C.~Cui, Z.~Fan, T.~Yang, P.~Lin, C.~Gong, and C.~Zhang, ``Domain
  adaptation-based transfer learning for drl control implementation of dc
  microgrids,'' \emph{IEEE Transactions on Industrial Electronics}, vol.~72,
  DOI
  10.1109/TIE.2025.3572981, no.~12, pp. 14\,344--14\,355, 2025.

\bibitem{TCSI.2023.3325590}
C.~Cui, Y.~Dong, X.~Dong, C.~Zhang, and A.~M. Y.~M. Ghias, ``Adaptive horizon
  seeking for generalized predictive control via deep reinforcement learning
  with application to dc/dc converters,'' \emph{IEEE Transactions on Circuits
  and Systems I: Regular Papers}, vol.~71,
  DOI
  10.1109/TCSI.2023.3325590, no.~5, pp. 2217--2228, 2024.

\bibitem{JESTPE.2022.3225264}
C.~Zhang, M.~Li, L.~Zhou, C.~Cui, and L.~Xu, ``A variable self-tuning horizon
  mechanism for generalized dynamic predictive control on dc/dc boost
  converters feeding cpls,'' \emph{IEEE Journal of Emerging and Selected Topics
  in Power Electronics}, vol.~11,
  DOI
  10.1109/JESTPE.2022.3225264, no.~2, pp. 1650--1660, 2023.

\bibitem{JESTPE.2023.3347515}
L.~Zhou, C.~Zhang, C.~Cui, P.~Lin, and X.~Dong, ``A drl-based parameter self
  configuration mechanism of nonsmooth control for autonomous dc microgrids
  feeding constant power loads,'' \emph{IEEE Journal of Emerging and Selected
  Topics in Power Electronics}, vol.~12,
  DOI
  10.1109/JESTPE.2023.3347515, no.~1, pp. 641--650, 2024.

\bibitem{TII.2024.3507191}
X.~Dong, C.~Zhang, H.~Chen, and W.~Zhang, ``A drl-based adaptive control design
  for a class of nonlinear systems with mismatched disturbances: From algorithm
  to application,'' \emph{IEEE Transactions on Industrial Informatics},
  vol.~21, DOI
  10.1109/TII.2024.3507191, no.~5, pp. 4126--4135, 2025.

\bibitem{TSTE.2023.3341632}
M.~Zhang, G.~Guo, S.~Magn{\'u}sson, R.~C.~N. Pilawa-Podgurski, and Q.~Xu,
  ``Data driven decentralized control of inverter based renewable energy
  sources using safe guaranteed multi-agent deep reinforcement learning,''
  \emph{IEEE Transactions on Sustainable Energy}, vol.~15,
  DOI
  10.1109/TSTE.2023.3341632, no.~2, pp. 1288--1299, 2024.

\bibitem{TPWRS.2023.3336614}
M.~Zhang, G.~Guo, T.~Zhao, and Q.~Xu, ``Dnn assisted projection based deep
  reinforcement learning for safe control of distribution grids,'' \emph{IEEE
  Transactions on Power Systems}, vol.~39,
  DOI
  10.1109/TPWRS.2023.3336614, no.~4, pp. 5687--5698, 2024.

\bibitem{TSG.2022.3228636}
P.~Chen, S.~Liu, X.~Wang, and I.~Kamwa, ``Physics-shielded multi-agent deep
  reinforcement learning for safe active voltage control with
  photovoltaic/battery energy storage systems,'' \emph{IEEE Transactions on
  Smart Grid}, vol.~14, DOI
  10.1109/TSG.2022.3228636, no.~4, pp. 2656--2667, 2023.

\bibitem{TPWRS.2024.3483994}
X.~Wan and M.~Sun, ``Adapsafe2: Prior-free safe-certified reinforcement
  learning for multi-area frequency control,'' \emph{IEEE Transactions on Power
  Systems}, vol.~40, DOI
  10.1109/TPWRS.2024.3483994, no.~3, pp. 2244--2257, 2025.

\bibitem{TASE.2025.3554431}
B.~Farzanegan and S.~Jagannathan, ``Explainable and safety aware deep
  reinforcement learning-based control of nonlinear discrete-time systems using
  neural network gradient decomposition,'' \emph{IEEE Transactions on
  Automation Science and Engineering}, vol.~22,
  DOI
  10.1109/TASE.2025.3554431, pp. 13\,556--13\,568, 2025.

\bibitem{TPWRS.2023.3326121}
T.~Wu, A.~Scaglione, and D.~Arnold, ``Constrained reinforcement learning for
  predictive control in real-time stochastic dynamic optimal power flow,''
  \emph{IEEE Transactions on Power Systems}, vol.~39,
  DOI
  10.1109/TPWRS.2023.3326121, no.~3, pp. 5077--5090, 2024.

\bibitem{TCNS.2021.3074218}
A.~Carron, K.~P. Wabersich, and M.~N. Zeilinger, ``Plug-and-play distributed
  safety verification for linear control systems with bounded uncertainties,''
  \emph{IEEE Transactions on Control of Network Systems}, vol.~8,
  DOI
  10.1109/TCNS.2021.3074218, no.~3, pp. 1501--1512, 2021.

\bibitem{TIE.2024.3363759}
Y.~Wan, Q.~Xu, and T.~Dragi{\v{c}}evi{\'c}, ``Safety-enhanced self-learning for
  optimal power converter control,'' \emph{IEEE Transactions on Industrial
  Electronics}, vol.~71, DOI
  10.1109/TIE.2024.3363759, no.~11, pp. 15\,229--15\,234, 2024.

\bibitem{MPCE.2023.000882}
H.~Shuai, B.~She, J.~Wang, and F.~Li, ``Safe reinforcement learning for
  grid-forming inverter based frequency regulation with stability guarantee,''
  \emph{Journal of Modern Power Systems and Clean Energy}, vol.~13,
  DOI
  10.35833/MPCE.2023.000882, no.~1, pp. 79--86, 2025.

\bibitem{TSG.2025.3616402}
M.~Eichelbeck, H.~Markgraf, and M.~Althoff, ``Commonpower: A framework for safe
  data-driven smart grid control,'' \emph{IEEE Transactions on Smart Grid},
  vol.~17, DOI
  10.1109/TSG.2025.3616402, no.~1, pp. 71--82, 2026.

\bibitem{TIA.2025.3572075}
X.~Li, Z.~Sun, X.~Zhang, Y.~Jiang, K.~Mao, and Y.~Yang, ``A robust artificial
  intelligence-empowered adaptive proportional-integral control for wireless
  power transfer systems,'' \emph{IEEE Transactions on Industry Applications},
  vol.~62, DOI
  10.1109/TIA.2025.3572075, no.~1, pp. 1433--1443, 2026.

\bibitem{cordova2023aggregate}
S.~C{\'o}rdova, C.~A. Ca{\~n}izares, {\'A}.~Lorca, and D.~E. Olivares,
  ``Aggregate modeling of thermostatically controlled loads for microgrid
  energy management systems,'' \emph{IEEE Transactions on Smart Grid}, vol.~14,
  no.~6, pp. 4169--4181, 2023.

\bibitem{wang2024nonconvex}
Z.-Y. Wang and H.-D. Chiang, ``On the nonconvex feasible region of optimal
  power flow: Theory, degree, and impacts,'' \emph{International Journal of
  Electrical Power \& Energy Systems}, vol. 161, p. 110167, 2024.

\bibitem{dominguez2023twin}
D.~Dominguez-Barbero, J.~Garcia-Gonzalez, and M.~A. Sanz-Bobi, ``Twin-delayed
  deep deterministic policy gradient algorithm for the energy management of
  microgrids,'' \emph{Engineering Applications of Artificial Intelligence},
  vol. 125, p. 106693, 2023.

\bibitem{asghar2025deep}
E.~Asghar, I.~Sengor, M.~Hill, and C.~Lynch, ``Deep q-network for intelligent
  energy management system in an energy community,'' \emph{Energy Strategy
  Reviews}, vol.~62, p. 101991, 2025.

\bibitem{li2023multi}
S.~Li, D.~Cao, W.~Hu, Q.~Huang, Z.~Chen, and F.~Blaabjerg, ``Multi-energy
  management of interconnected multi-microgrid system using multi-agent deep
  reinforcement learning,'' \emph{Journal of Modern Power Systems and Clean
  Energy}, vol.~11, no.~5, pp. 1606--1617, 2023.

\bibitem{talaat2023artificial}
M.~Talaat, M.~Elkholy, A.~Alblawi, and T.~Said, ``Artificial intelligence
  applications for microgrids integration and management of hybrid renewable
  energy sources,'' \emph{Artificial Intelligence Review}, vol.~56, no.~9, pp.
  10\,557--10\,611, 2023.

\bibitem{zhang2023multi}
B.~Zhang, W.~Hu, A.~M. Ghias, X.~Xu, and Z.~Chen, ``Multi-agent deep
  reinforcement learning based distributed control architecture for
  interconnected multi-energy microgrid energy management and optimization,''
  \emph{Energy Conversion and Management}, vol. 277, p. 116647, 2023.

\bibitem{r2024machine}
A.~R.~Singh, R.~S. Kumar, M.~Bajaj, C.~B. Khadse, and I.~Zaitsev, ``Machine
  learning-based energy management and power forecasting in grid-connected
  microgrids with multiple distributed energy sources,'' \emph{Scientific
  Reports}, vol.~14, no.~1, p. 19207, 2024.

\bibitem{ozcanli2022islanding}
A.~K. Ozcanli and M.~Baysal, ``Islanding detection in microgrid using deep
  learning based on 1d cnn and cnn-lstm networks,'' \emph{Sustainable Energy,
  Grids and Networks}, vol.~32, p. 100839, 2022.

\bibitem{li2023deep}
Y.~Li, C.~Yu, M.~Shahidehpour, T.~Yang, Z.~Zeng, and T.~Chai, ``Deep
  reinforcement learning for smart grid operations: Algorithms, applications,
  and prospects,'' \emph{Proc. IEEE}, vol. 111, no.~9, pp. 1055--1096, 2023.

\bibitem{wang2024live}
X.~Wang, S.~Li, and M.~Iqbal, ``Live power generation predictions via ai-driven
  resilient systems in smart microgrids,'' \emph{IEEE Transactions on Consumer
  Electronics}, vol.~70, no.~1, pp. 3875--3884, 2024.

\bibitem{mansour2024wasserstein}
S.~H. Mansour, S.~M. Azzam, H.~M. Hasanien, M.~Tostado-Veliz, A.~Alkuhayli, and
  F.~Jurado, ``Wasserstein generative adversarial networks-based photovoltaic
  uncertainty in a smart home energy management system including battery
  storage devices,'' \emph{Energy}, vol. 306, p. 132412, 2024.

\bibitem{aly2022hybrid}
H.~H. Aly, ``A hybrid optimized model of adaptive neuro-fuzzy inference system,
  recurrent kalman filter and neuro-wavelet for wind power forecasting driven
  by dfig,'' \emph{Energy}, vol. 239, p. 122367, 2022.

\bibitem{jeyaraj2023deep}
P.~R. Jeyaraj, S.~P. Asokan, A.~C. Kathiresan, and E.~R.~S. Nadar, ``Deep
  reinforcement learning-based network for optimized power flow in islanded dc
  microgrid,'' \emph{Electrical engineering}, vol. 105, no.~5, pp. 2805--2816,
  2023.

\bibitem{wang2023towards}
Y.~Wang, D.~Qiu, F.~Teng, and G.~Strbac, ``Towards microgrid resilience
  enhancement via mobile power sources and repair crews: A multi-agent
  reinforcement learning approach,'' \emph{IEEE transactions on power systems},
  vol.~39, no.~1, pp. 1329--1345, 2023.

\bibitem{ahsan2025multi}
S.~M. Ahsan, N.~Gholizadeh, and P.~Musilek, ``Multi-agent systems in networked
  microgrids: Reinforcement learning and strategic pricing mechanisms,''
  \emph{Renewable Energy}, vol. 254, p. 123678, 2025.

\bibitem{sayal2024ai}
A.~Sayal, C.~N, J.~Jha, and N.~Allagari, ``Ai-based predictive maintenance
  strategies for improving the reliability of green power systems,'' in
  \emph{Digital Technologies to Implement the UN Sustainable Development
  Goals}, pp. 19--46.\hskip 1em plus 0.5em minus 0.4em\relax Springer, 2024.

\bibitem{omo2026gmdh}
E.~Omo-Ikerodah, A.~Etminan, and M.~Jamil, ``Gmdh-enhanced temporal
  convolutional network for short-term wind forecasting and microgrid
  operation,'' \emph{Energy Conversion and Management}, vol. 357, p. 121404,
  2026.

\bibitem{mohamed2022dynamic}
M.~Mohamed, F.~E. Mahmood, M.~A. Abd, A.~Chandra, and B.~Singh, ``Dynamic
  forecasting of solar energy microgrid systems using feature engineering,''
  \emph{IEEE Transactions on Industry Applications}, vol.~58, no.~6, pp.
  7857--7869, 2022.

\bibitem{zahid2026ai}
M.~Zahid, H.~M. Munir, M.~Adeel, F.~S. Alromithy, M.~R. Altimania, and
  I.~Zaitsev, ``Ai-driven optimization techniques for power quality improvement
  in microgrids: Trends, techniques, and future directions,'' \emph{Energy
  Science \& Engineering}, vol.~14, no.~1, pp. 583--610, 2026.

\bibitem{cano2024integrating}
A.~Cano, P.~Ar{\'e}valo, D.~Benavides, and F.~Jurado, ``Integrating discrete
  wavelet transform with neural networks and machine learning for fault
  detection in microgrids,'' \emph{International Journal of Electrical Power \&
  Energy Systems}, vol. 155, p. 109616, 2024.

\bibitem{sitharthan2023smart}
R.~Sitharthan, S.~Vimal, A.~Verma, M.~Karthikeyan, S.~S. Dhanabalan,
  N.~Prabaharan, M.~Rajesh, and T.~Eswaran, ``Smart microgrid with the internet
  of things for adequate energy management and analysis,'' \emph{Computers and
  Electrical Engineering}, vol. 106, p. 108556, 2023.

\bibitem{veerasamy2023blockchain}
V.~Veerasamy, L.~P. M.~I. Sampath, S.~Singh, H.~D. Nguyen, and H.~B. Gooi,
  ``Blockchain-based decentralized frequency control of microgrids using
  federated learning fractional-order recurrent neural network,'' \emph{IEEE
  transactions on smart grid}, vol.~15, no.~1, pp. 1089--1102, 2023.

\bibitem{veerasamy2024blockchain}
V.~Veerasamy, Z.~Hu, H.~Qiu, S.~Murshid, H.~B. Gooi, and H.~D. Nguyen,
  ``Blockchain-enabled peer-to-peer energy trading and resilient control of
  microgrids,'' \emph{Applied energy}, vol. 353, p. 122107, 2024.

\bibitem{fan2023energy}
X.~Fan and Y.~Li, ``Energy management of renewable based power grids using
  artificial intelligence: Digital twin of renewables,'' \emph{Solar Energy},
  vol. 262, p. 111867, 2023.

\bibitem{hadi2025artificial}
M.~Hadi, E.~Elbouchikhi, Z.~Zhou, A.~Saim, M.~Shafie-Khah, P.~Siano,
  H.~Rahbarimagham, and P.~M. Colom, ``Artificial intelligence for microgrids
  design, control, and maintenance: A comprehensive review and prospects,''
  \emph{Energy Conversion and Management: X}, vol.~27, p. 101056, 2025.

\bibitem{pervez2025computationally}
I.~Pervez, C.~Antoniadis, and H.~Ghazzai, ``Computationally efficient model
  predictive control for enhanced primary-level management of standalone
  microgrids with hybrid storage systems,'' \emph{IEEE Transactions on
  Sustainable Energy}, 2025.

\bibitem{yan2020real}
Z.~Yan and Y.~Xu, ``Real-time optimal power flow: A lagrangian based deep
  reinforcement learning approach,'' \emph{IEEE Transactions on Power Systems},
  vol.~35, no.~4, pp. 3270--3273, 2020.

\bibitem{yan2023real}
Z.~Yan and Y.~Xu, ``Real-time optimal power flow with linguistic stipulations:
  Integrating gpt-agent and deep reinforcement learning,'' \emph{IEEE
  Transactions on Power Systems}, vol.~39, no.~2, pp. 4747--4750, 2023.

\bibitem{9108633}
Z.~Yan and Y.~Xu, ``A multi-agent deep reinforcement learning method for
  cooperative load frequency control of a multi-area power system,'' \emph{IEEE
  Transactions on Power Systems}, vol.~35,
  DOI
  10.1109/TPWRS.2020.2999890, no.~6, pp. 4599--4608, 2020.

\bibitem{davoudkhani2024robust}
I.~F. Davoudkhani, P.~Zare, A.~Y. Abdelaziz, M.~Bajaj, and M.~B. Tuka, ``Robust
  load-frequency control of islanded urban microgrid using 1pd-3dof-pid
  controller including mobile ev energy storage,'' \emph{Scientific Reports},
  vol.~14, no.~1, p. 13962, 2024.

\bibitem{hua2022applications}
W.~Hua, Y.~Chen, M.~Qadrdan, J.~Jiang, H.~Sun, and J.~Wu, ``Applications of
  blockchain and artificial intelligence technologies for enabling prosumers in
  smart grids: A review,'' \emph{Renewable and Sustainable Energy Reviews},
  vol. 161, p. 112308, 2022.

\bibitem{ren2024transferable}
Y.~Ren, H.~Zhang, W.~Yang, M.~Li, J.~Zhang, and H.~Li, ``Transferable
  adversarial attack against deep reinforcement learning-based smart grid
  dynamic pricing system,'' \emph{IEEE Transactions on Industrial Informatics},
  vol.~20, no.~6, pp. 9015--9025, 2024.

\bibitem{boza2021artificial}
P.~Boza and T.~Evgeniou, ``Artificial intelligence to support the integration
  of variable renewable energy sources to the power system,'' \emph{Applied
  Energy}, vol. 290, p. 116754, 2021.

\bibitem{ahmad2018distribution}
F.~Ahmad, A.~Rasool, E.~Ozsoy, R.~Sekar, A.~Sabanovic, and M.~Elita{\c{s}},
  ``Distribution system state estimation-a step towards smart grid,''
  \emph{Renewable and Sustainable Energy Reviews}, vol.~81, pp. 2659--2671,
  2018.

\bibitem{khashei2025mean}
M.~Khashei, M.~Ahmadi, and F.~Chahkoutahi, ``A mean weighted squared
  error-based neural classifier for intelligent pattern recognition in smart
  grids,'' \emph{International Journal of Electrical Power \& Energy Systems},
  vol. 170, p. 110972, 2025.

\bibitem{qiu2024artificial}
D.~Qiu, G.~Strbac, Y.~Wang, Y.~Ye, J.~Wang, P.~Pinson, V.~Silva, and F.~Teng,
  ``Artificial intelligence for microgrid resilience: a data-driven and
  model-free approach,'' \emph{IEEE Power and Energy Magazine}, vol.~22, no.~6,
  pp. 18--27, 2024.

\bibitem{yan2020multi}
Z.~Yan and Y.~Xu, ``A multi-agent deep reinforcement learning method for
  cooperative load frequency control of a multi-area power system,'' \emph{IEEE
  Transactions on Power Systems}, vol.~35, no.~6, pp. 4599--4608, 2020.

\bibitem{rocha2021artificial}
H.~R. Rocha, I.~H. Honorato, R.~Fiorotti, W.~C. Celeste, L.~J. Silvestre, and
  J.~A. Silva, ``An artificial intelligence based scheduling algorithm for
  demand-side energy management in smart homes,'' \emph{Applied Energy}, vol.
  282, p. 116145, 2021.

\bibitem{zheng2021vulnerability}
Y.~Zheng, Z.~Yan, K.~Chen, J.~Sun, Y.~Xu, and Y.~Liu, ``Vulnerability
  assessment of deep reinforcement learning models for power system topology
  optimization,'' \emph{IEEE Transactions on Smart Grid}, vol.~12, no.~4, pp.
  3613--3623, 2021.

\bibitem{liu2025noise}
Y.~Liu and J.~Ma, ``Noise-trained deep learning-based distribution system state
  estimation considering the penetration of distributed energy resources,''
  \emph{IEEE Transactions on Smart Grid}, vol.~16, no.~3, pp. 2292--2303, 2025.

\bibitem{barja2021artificial}
S.~Barja-Martinez, M.~Arag{\"u}{\'e}s-Pe{\~n}alba, {\'I}.~Munn{\'e}-Collado,
  P.~Lloret-Gallego, E.~Bullich-Massagu{\'e}, and R.~Villafafila-Robles,
  ``Artificial intelligence techniques for enabling big data services in
  distribution networks: A review,'' \emph{Renewable and Sustainable Energy
  Reviews}, vol. 150, p. 111459, 2021.

\bibitem{ge2025autonomous}
L.~Ge, J.~Li, L.~Hou, and J.~Lai, ``Autonomous voltage regulation for smart
  distribution network with high-proportion pvs: A graph meta-reinforcement
  learning approach,'' \emph{IEEE Transactions on Sustainable Energy}, 2025.

\bibitem{dindar2026privacy}
\BIBentryALTinterwordspacing
D.~Burak, B.~S. Can, K.~C. Huseyin, and H.~Veit, ``Privacy-preserving
  utilization of distribution system flexibility for enhanced tso-dso
  interoperability: A novel machine learning-based optimal power flow
  approach,'' \emph{Applied Energy}, vol. 414,
  DOI
  https://doi.org/10.1016/j.apenergy.2026.127848, p. 127848, 2026. [Online].
  Available:
  \arxivurl{https://www.sciencedirect.com/science/article/pii/S0306261926005003}
\BIBentrySTDinterwordspacing

\bibitem{steven2026machine}
R.~Steven, O.~V. Klymenko, and M.~Short, ``Machine learning-accelerated
  distributed optimisation methods for optimal power flow: A review,''
  \emph{Renewable and Sustainable Energy Reviews}, vol. 226, p. 116190, 2026.

\bibitem{shang2021achieving}
Y.~Shang, Y.~Shang, H.~Yu, Z.~Shao, and L.~Jian, ``Achieving efficient and
  adaptable dispatching for vehicle-to-grid using distributed edge computing
  and attention-based lstm,'' \emph{IEEE Transactions on Industrial
  Informatics}, vol.~18, no.~10, pp. 6915--6926, 2021.

\bibitem{xiong2022two}
L.~Xiong, Y.~Tang, S.~Mao, H.~Liu, K.~Meng, Z.~Dong, and F.~Qian, ``A two-level
  energy management strategy for multi-microgrid systems with interval
  prediction and reinforcement learning,'' \emph{IEEE Transactions on Circuits
  and Systems I: Regular Papers}, vol.~69, no.~4, pp. 1788--1799, 2022.

\bibitem{baldwin2002power}
T.~L. Baldwin, L.~Mili, M.~B. Boisen, and R.~Adapa, ``Power system
  observability with minimal phasor measurement placement,'' \emph{IEEE
  Transactions on Power systems}, vol.~8, no.~2, pp. 707--715, 2002.

\bibitem{jeyaraj2022optimum}
P.~R. Jeyaraj, S.~P. Asokan, and A.~C. Karthiresan, ``Optimum power flow in dc
  microgrid employing bayesian regularized deep neural network,''
  \emph{Electric Power Systems Research}, vol. 205, p. 107730, 2022.

\bibitem{karimi2016photovoltaic}
M.~Karimi, H.~Mokhlis, K.~Naidu, S.~Uddin, and A.~A. Bakar, ``Photovoltaic
  penetration issues and impacts in distribution network--a review,''
  \emph{Renewable and sustainable energy reviews}, vol.~53, pp. 594--605, 2016.

\bibitem{shi2023coordinated}
W.~Shi, D.~Zhang, X.~Han, X.~Wang, T.~Pu, and W.~Chen, ``Coordinated operation
  of active distribution network, networked microgrids, and electric vehicle: A
  multi-agent ppo optimization method,'' \emph{CSEE Journal of Power and Energy
  Systems}, 2023.

\bibitem{cao2023physics}
D.~Cao, J.~Zhao, J.~Hu, Y.~Pei, Q.~Huang, Z.~Chen, and W.~Hu,
  ``Physics-informed graphical representation-enabled deep reinforcement
  learning for robust distribution system voltage control,'' \emph{IEEE
  Transactions on Smart Grid}, vol.~15, no.~1, pp. 233--246, 2023.

\bibitem{cunha2025unlocking}
L.~V. Cunha, A.~vd~Veen, B.~Espinosa, D.~Diran, D.~Ruoff, E.~Karangelos,
  E.~Sarmas, H.~Madsen, J.~Chaves, K.~ad~Horst \emph{et~al.}, ``Unlocking the
  potential of ai and generative ai in european smart grids--a strategic
  position paper and guide for action,'' \emph{European Commission}, 2025.

\bibitem{rahman2025us}
T.~Rahman, ``The us approach to cybersecurity in the energy sector,'' in
  \emph{The Palgrave Handbook of Cybersecurity, Technologies and Energy
  Transitions}, pp. 1--52.\hskip 1em plus 0.5em minus 0.4em\relax Springer,
  2025.

\bibitem{sarkar2023study}
S.~Sarkar, ``A study on cybersecurity standards for power systems,'' in
  \emph{Power Systems Cybersecurity: Methods, Concepts, and Best Practices},
  pp. 429--450.\hskip 1em plus 0.5em minus 0.4em\relax Springer, 2023.

\bibitem{badr2023privacy}
M.~M. Badr, M.~M. Mahmoud, Y.~Fang, M.~Abdulaal, A.~J. Aljohani, W.~Alasmary,
  and M.~I. Ibrahem, ``Privacy-preserving and communication-efficient energy
  prediction scheme based on federated learning for smart grids,'' \emph{IEEE
  Internet of Things Journal}, vol.~10, no.~9, pp. 7719--7736, 2023.

\bibitem{jorgensen2025impact}
B.~N. J{\o}rgensen, S.~S. Gunasekaran, and Z.~G. Ma, ``Impact of eu laws on ai
  adoption in smart grids: A review of regulatory barriers, technological
  challenges, and stakeholder benefits,'' \emph{Energies}, vol.~18, no.~12, p.
  3002, 2025.

\bibitem{lian2025secure}
Z.~Lian, H.~Yu, Y.~Chen, M.~Shahidehpour, and Q.~Zhou, ``Secure microgrid
  clusters under integrated satellite-terrestrial networks: A differential
  privacy-preserving control strategy,'' \emph{IEEE Transactions on Industrial
  Electronics}, 2025.

\bibitem{UL3115_AI_Safety}
\BIBentryALTinterwordspacing
{UL Solutions}, ``{UL 3115 AI Safety Certification Services for AI-Enabled
  Products},'' 2026, accessed: 2026-05-24. [Online]. Available:
  \arxivurl{https://www.ul.com/services/ul-3115-ai-safety-certification-services-ai-enabled-products}
\BIBentrySTDinterwordspacing

\end{thebibliography}
\end{document}